\documentclass[12pt]{article}
\pdfoutput=1
\usepackage[english]{babel}
\usepackage[T1]{fontenc}
\usepackage{amsmath,amsfonts,color,braket,cite,graphicx,color}
\usepackage{ulem}
\newcommand{\no}{\nonumber }
\newcommand{\be}{\begin{equation}}
\newcommand{\ee}{\end{equation}}
\newcommand{\ba}{\begin{aligned}}
\newcommand{\ea}{\end{aligned}}
\newcommand{\been}{\begin{eqnarray*}}
\newcommand{\een}{\end{eqnarray*}}
\newcommand{\beg}{\begin{equation}\begin{aligned}}
\newcommand{\eng}{\end{aligned}\end{equation}}

\begin{document}

\begin{titlepage}
\begin{flushright}
TIT/HEP-676\\
NORDITA 2019-092 \\
October,  2019
\end{flushright}
\vspace{0.5cm}
\begin{center}
{\Large \bf
TBA equations for the Schr\"odinger equation with a regular singularity
}
\lineskip .75em
\vskip 2.5cm

{\large  Katsushi Ito$^{a,}$
, Hongfei Shu$^{b,a,}$
}
\vskip 2.5em
 {\normalsize\it $^{a}$Department of Physics, Tokyo Institute of Technology
Tokyo, 152-8551, Japan,\\
$^{b}$Nordita, KTH Royal Institute of Technology and Stockholm University\\
			Roslagstullsbacken 23, SE-106 91 Stockholm, Sweden}
\vskip 3.0em
\end{center}
\begin{abstract}
We derive the Thermodynamic Bethe Ansatz (TBA) equations for the Schr\"odinger equation with an arbitrary polynomial potential and a regular singular (simple and double pole) term. The TBA equations provide a non-trivial generalization of the ODE/IM correspondence and also give a solution for the Riemann-Hilbert problem in the exact WKB method.  We study the TBA equations in detail for the linear and the harmonic oscillator potentials together with inverse and centrifugal terms. 
As an application, we also compute numerically the Voros spectrum for these potentials using the Bohr-Sommerfeld quantization condition.

  \end{abstract}
\end{titlepage}

\baselineskip=0.7cm



\section{Introduction}
The exact WKB method for stationary one-dimensional Schr\"odinger equation has been studied extensively for many years\cite{bpv79, voros79, voros83, ddp93, ddp97, dp99}. The exact WKB periods (or the quantum periods) are asymptotic series in the Planck constant and therefore must be treated  by Borel and lateral Laplace transformations.  
Resurgence  helps us to understand the relation between perturbative and non-perturbative effects to the periods.  In particular, their classical limit and discontinuity structure (the Stokes phenomena) determine the quantum periods completely.  This analytic bootstrap  program (or the Riemann-Hilbert problem) has been worked out for cubic and quartic potentials \cite{voros83}. 

 The Stokes phenomena for the solutions of the Schr\"odinger equation in the complex plane leads to the functional relations among the Wronskians of the solutions. Incorporating the basis of solutions around the origin, these functional relations are identified with the functional equations  of the quantum integrable models such as T-Q relations, T-system and Y-system. This phenomena is known as the ODE/IM correspondence \cite{Dorey:1998pt, Bazhanov:1998wj, Dorey:1999uk, Dorey:2007zx}, which has been generalized to higher-order differential equations\cite{Dorey:1999pv, Suzuki:1999hu, Dorey:2000ma, Suzuki:2000gi, Dorey:2006an, Sun:2012xw,Masoero:2015lga,Masoero:2015rcz}. This relation has been studied mainly in the 
 case of monic potential and the spectral problem has been solved by the NonLinear Integral Equations\footnote{See also \cite{Lukyanov:2010rn, Dorey:2012bx, Ito:2013aea, Adamopoulou:2014fca, Ito:2015nla, Ito:2016qzt, Ito:2018wgj} for the ODE/IM correspondence associated with affine Lie algebras.}.

Similar Riemann-Hilbert problems also appear in the study of the BPS spectrum in ${\cal N}=2$ supersymmetric gauge theories and the minimal surface in AdS background \cite{Kontsevich:2008fj, Gaiotto:2008cd, Gaiotto:2009hg, Alday:2009yn, Alday:2009dv, Alday:2010vh, Hatsuda:2010cc, Cavaglia:2010nm, Maldacena:2010kp, Gao:2013dza, Gaiotto:2014bza,Toledo10, Toledo16}.
 Inspired by these works, 
 the discontinuity formula for a general polynomial potential has been reformulated in the form
 of the thermodynamic Bethe ansatz (TBA) equations \cite{Ito:2018eon}, where exponential of the quantum periods over specific 
 one-cycles are identified as the Y-functions and the Planck constant plays a role of the spectral parameter. 
From the wall-crossing of the TBA equations, 
one can study the quantum periods for a polynomial potential with generic complex coefficients. 
 The TBA equations together with the exact quantization condition provide an efficient numerical method to
 solve the spectral problem of quantum mechanics.

 An interesting generalization is to add inverse and centrifugal terms to the potential.
 For a monomial potential with a centrifugal term, the ODE/IM correspondence has been studied in \cite{Dorey:1998pt, Bazhanov:1998wj, Dorey:1999uk, Dorey:2007zx, Suzuki:2015vwa}.
 The centrifugal term induces monodromy around the origin and the TBA system is extended to the $D$-type from the A-type, which is particularly important in understanding of the classification of integrable models. Moreover, by adding the centrifugal term, the Schr\"odinger equation can be regarded as the quantum Seiberg-Witten curve of the $(A_1, D_n)$ Argyres-Douglas theory \cite{Ito:2017ypt}, which helps us to realize the BPS spectrum of the strongly interacting SCFs.  In this work, we will study the Y-system and the TBA equations 
 for a general polynomial potential with a centrifugal potential.
 We will begin with the case where  all the turning points are located along the real axis.
 It becomes a non-trivial problem to choose appropriate one-cycle on the WKB curve
 for the Y-functions since in the classical limit $\hbar\rightarrow 0$ some turning points degenerate and the next-order corrections to the WKB periods become divergent. In order to regularize classical turning points, we add an inverse potential. 
For the inverse potential, the structure of the WKB lines has been studied in \cite{dr04, Iwaki-Nakanishi-1, Iwaki-Nakanishi-2}.
We will show that there appear new Y-functions  associated with non-trivial one-cycles on the Riemann surface  
and the TBA system is enhanced to the D-type.
By applying the wall-crossing, we expect to relate the present TBA to the system
in \cite{Dorey:1998pt, Bazhanov:1998wj}. But we will not discuss this issue in detail in this paper.

 This paper is organized as follows. In section 2, we derive the Y-system from the Schr\"odinger equation with arbitrary polynomial potential incorporated with simple and double pole potentials. 
 In section 3, we evaluate the asymptotic behavior of the Y-function and provide a closed TBA system. Some general properties of the TBA system are also presented. In section 4, we first re-derive the TBA equations from the discontinuity formula of WKB periods, which implies that our TBA equations govern the exact WKB periods. We then confirm numerically  the TBA equations by comparing the expansion of the TBA equations at large $\theta$ and the higher-order corrections of WKB periods. As an application, we compute the Voros spectrum numerically using the Bohr-Sommerfeld quantization condition.
  Finally, we present conclusions and discussions in section 6. In appendix A, we derive the T-Q relation and the Bethe ansatz equation from the Schr\"odinger equation. In appendix B, some details of the wall crossing of the TBA equations are presented. In appendix C, we discuss the relations among the present ODE and other types of ODEs under a change of variable.
This provides a further consistency check of the correspondence between the ODEs and the TBA systems.


\section{Y-system from the Schr\"odinger-type equation}
We consider the second order ODE of the Schr\"odinger type with a polynomial potential incorporated with a simple and a double pole terms:
\begin{align}\label{eq:Schrodinger-eq-zb}
\left(-\frac{d^2}{dz^2}+z^{r+1}+\sum_{a=1}^{r+2}b_{a}z^{r+1-a}+\frac{l(l+1)}{z^2}\right){\psi(z, b_a,l)}=0,
\end{align}
where $z$ is a complex variable, $r$ a non-negative integer, $l$ a real parameter,  and $b_a$, $a=1,\cdots,r+2$, complex parameters. This ODE possesses an irregular singularity at infinity and a regular singularity (simple and double pole) at origin. The double pole leads to a non-trivial monodromy around the origin, which is the main difference compared with the setup in \cite{Ito:2018eon}. 
We will study the functional relations satisfied the Stokes multipliers of the solution to the ODE (\ref{eq:Schrodinger-eq-zb}), following the approach in \cite{Bazhanov:1998wj, Dorey:1999uk} where the parameters $b_a$, $a=1,2,\cdots, r, r+2$ are zero. 

We first consider the behavior of the solution around $z=\infty$. One finds the fastest decaying solution of (\ref{eq:Schrodinger-eq-zb}) at infinity along the positive real axis behaves as \cite{Sibuya}
\begin{align}
y(z, b_a)\sim\frac{1}{\sqrt{2i}}z^{n_{r}}\exp\Big(-\frac{2}{r+3}z^{\frac{r+3}{2}}\Big),\label{eq:Decay-solution-S0}
\end{align}
where $n_r$ are defined by
\begin{align}
n_{r}&=\begin{cases}
-\frac{r+1}{4} & \quad r:\mbox{even}\\
-\frac{r+1}{4}-B_{\frac{r+3}{2}} & \quad r:\mbox{odd}
\end{cases}.
\end{align}
$B_m$ is defined by the following expansion:
\be
\Big(1+\sum_{a=1}^{r+2}b_az^{-a}+\frac{l(l+1)}{z^{r+3}}\Big)^{1/2}=:1+\sum_{m=1}^{\infty}B_{m}z^{-m}.
\ee
The equation (\ref{eq:Schrodinger-eq-zb}) is invariant under the Symanzik rotation $(z, b_a, l)\to (\omega z, \omega^{a}b_a, l)$:
\begin{align}
&\omega^{-2}\left(-\frac{d^2}{dz^2}+\omega^{r+3}z^{r+1}+\omega^{r+3}\sum_{a=1}^{r+2}b_{a}z^{r+1-a}+\frac{l(l+1)}{z^2}\right)\psi(\omega z, \omega^{a}b_{a},l)=0
\end{align}
with $\omega=e^{\frac{2\pi i}{r+3}}$. 
By using the Symanzik rotation, we can construct a new solution $y_k$ ($k\in {\mathbb Z}$) from $y$:
\be
\label{eq:yk-solu}
y_{k}(z, b_{a}, l)=\omega^{\frac{k}{2}}y(\omega^{-k}z,\omega^{-ak}b_{a}, l).
\ee
Since $l$ does not change under the Symanzik rotation, we will omit the argument $l$ in the solution.
The solution $y_k$ is subdominant in the sector ${\cal S}_k$, which is  defined by 
\be
{\cal S}_k=\left\{ z \in \mathbb{C}:  \left|\mbox{arg}(z)-\frac{2k\pi}{r+3}\right|<\frac{\pi}{r+3} \right\}.
\ee
But $y_{k\pm 1}$ is a dominant solution in the sector ${\cal S}_k$. Then we can choose $\{ y_{k}$, $y_{k+1}\} $ as a basis of the solutions of (\ref{eq:Schrodinger-eq-zb}). We define the Wronskian of $y_{k_1}$ and $y_{k_2}$ by
\begin{align}
W_{k_1,k_2}(b_a)\equiv W[y_{k_1}(z,b_a), y_{k_2}(z,b_a)]=y_{k_1}(z,b_a)\partial_zy_{k_2}(z,b_a)-y_{k_2}(z,b_a)\partial_zy_{k_1}(z,b_a).
\end{align}
One finds $W_{k_1,k_2}^{[2]}(b_a)=W_{k_1+1, k_2+1}(b_a)$, where we have introduced the $j$-shift of the phase in the arguments of a function by 
\begin{align}
f^{[j]}(z,b_{a})&:=f(\omega^{-j/2}z,\omega^{-ja/2}b_{a})
\label{eq:shift1}
\end{align}
for an integer $j$. In our normalization (\ref{eq:Decay-solution-S0}), the Wronskian of $y_k$ and $y_{k+1}$ is evaluated as
\begin{align}\label{eq:normalization}
W_{k,k+1}(b_a)&=\begin{cases}
1 & \quad r:\mbox{even}\\
\omega^{(-1)^{k}B_{\frac{r+3}{2}}} & \quad r:\mbox{odd}
\end{cases}.
\end{align}
When we  choose $\{ y_0, y_1\}$ as the basis, $y_k$ is expanded as
\be
\label{eq:yk-y0y1}
y_{k}(x,b_a)=\frac{W_{k,1}}{W_{0,1}}y_{0}(x,b_a)+\frac{W_{0,k}}{W_{0,1}}y_{1}(x,b_a).
\ee
The ratios of the Wronskians  are called the Stokes multipliers which relates the solutions in different sectors.

\subsection{Monodromy around origin} \label{sec:monodromy}
Next, we consider the monodromy of the solutions around the origin.
Note that the sectors ${\cal S}_k$ provide a multi-cover for the complex plane, where consecutive $r+3$ sectors cover the single complex plane. Since $y_{j+r+3}(z)$ and $y_j(ze^{-2\pi i})$ are the decaying solutions at infinity in the same sector, $y_{j+r+3}(z)\propto y_j(ze^{-2\pi i})$.  From the definition (\ref{eq:yk-solu}), we find 
\be
\label{eq:yr+3-y0}
y_{r+3}(z,b_{a})=\omega^{\frac{r+3}{2}}y_{0}(e^{-2\pi i}z,b_{a}).
\ee
By acting the Symanzik rotation on both sides, one finds
\be
\label{eq:yr+4-y1}
y_{r+4}(z,b_{a})=\omega^{\frac{r+3}{2}}y_{1}(e^{-2\pi i}z,b_{a}).
\ee
Since $y_{r+3}$ and  $y_{r+4}$ are expanded in the basis $\{y_0,y_1\}$,
$y_0(e^{-2\pi i}z)$ and $y_1(e^{-2\pi i}z)$ are also expanded in the same basis.
Let us introduce the monodromy matrix $\Omega(b_a, l)$ by
\be
\left(\begin{array}{c}
y_{1}\\
y_{0}
\end{array}\right)(e^{-2\pi i}z)=\Omega(b_{a}, l)\left(\begin{array}{c}
y_{1}\\
y_{0}
\end{array}\right)(z).
\ee
Then 
the relations (\ref{eq:yr+3-y0}) and (\ref{eq:yr+4-y1}) imply
\be
\left(\begin{array}{c}
y_{r+4}\\
y_{r+3}
\end{array}\right)(z)=\omega^{\frac{r+3}{2}}\Omega(b_{a}, l)\left(\begin{array}{c}
y_{1}\\
y_{0}
\end{array}\right)(z).
\ee
From
(\ref{eq:yk-y0y1}), one finds the monodromy matrix $\Omega(b_a, l)$ can be expressed in terms of the Wronskians:
\be
\Omega(b_{a}, l)=\omega^{-\frac{r+3}{2}}\left(\begin{array}{cc}
\frac{W_{0,r+4}}{W_{0,1}} & \frac{W_{r+4,1}}{W_{0,1}}\\
\frac{W_{0,r+3}}{W_{0,1}} & \frac{W_{r+3,1}}{W_{0,1}}
\end{array}\right),
\ee
whose trace is evaluated as
\be
\label{eq:trace-mono-Wron}
{\rm Tr}[\Omega(b_{a}, l)]=\omega^{-\frac{r+3}{2}}\Big(\frac{W_{0,r+4}}{W_{r+3,r+4}}+\frac{W_{r+3,1}}{W_{0,1}}\Big).
\ee

From  the power series solutions of  the Schr\"odinger-type equation (\ref{eq:Schrodinger-eq-zb}) around the origin, we can evaluate the trace of monodromy matrix. In the small $z$ region, there are power series solutions of the form
\beg\label{eq:basis-z=0}
\psi_{+}(z,b_{a},l)&=z^{l+1}+{\cal O}(z^{l+3}),\quad\psi_{-}(z,b_{a},l)=z^{-l}+{\cal O}(z^{-l+2}).
\eng
$\{\psi_+,\psi_-\}$ forms a basis of the solutions around $z=0$. The monodromy matrix ${\cal M}(b_{a},l)$ for the basis $(\psi_+,\psi_-)$ is defined by
 \be
\left(\begin{array}{c}
\psi_{+}\\
\psi_{-}
\end{array}\right)(e^{-2\pi i}z,b_{a})={\cal M}(b_{a},l)\left(\begin{array}{c}
\psi_{+}\\
\psi_{-}
\end{array}\right)(z), \quad 
{\cal M}(b_a,l)=
\left(
\begin{array}{cc}
e^{-2\pi (l+1)i} & 0\\
0 & e^{2\pi l i}
\end{array}
\right),
\ee
which is diagonalized in this basis.
Their Wronskian of $\psi_+$ and $\psi_-$ is evaluated as
\beg
W[\psi_+,\psi_-]=-(2l+1).
\eng
In terms of the new basis $\{ \psi_+, \psi_-\}$, $y_0$ and $y_1$ can be expanded as
\beg
\left(\begin{array}{c}
y_{1}\\
y_{0}
\end{array}\right)(z)&={\cal Q}(b_{a})\left(\begin{array}{c}
\psi_{+}\\
\psi_{-}
\end{array}\right)(z).
\eng
The matrix ${\cal Q}(b_a)$ defines the Q-functions of certain integrable models.
The functional relations satisfied by ${\cal Q}(b_a)$ can be found in the appendix \ref{sec:TQ-BAE}. The monodromy matrix $\Omega(b_a,l)$ is diagonalized by ${\cal Q}(b_a,l)$ as 
\be
\Omega(b_{a}, l)={\cal Q}(b_{a}){\cal M}{\cal Q}(b_{a})^{-1}.
\ee
Then the trace of the monodormy matrix is given by
\be
\label{eq:trcae-monodromy}
{\rm Tr}[\Omega(b_a, l)]={\rm Tr}[{\cal M}(b_a,l)]=2\cos(2\pi l),
\ee
which depends only on $l$.

\subsection{Truncated Y-system}
From the Q-functions ${\cal Q}(b_a)$, we can derive the Baxter's T-Q relations  and the Bethe ansatz equations, which are explained in Appendix \ref{sec:TQ-BAE}. In this subsection, we focus on the Y-system.
Inspired by the construction of the Y-system in \cite{Alday:2010vh} for the Hitchin system associated with $AdS_3$, we introduce the set of functions of $b_a$ by \cite{Ito:2018eon}
\be
\ba
{\cal Y}_{2j}(b_a)&:=\frac{W_{-j,j}W_{-j-1,j+1}}{W_{-j-1,-j}W_{j,j+1}}(b_a),\\
{\cal Y}_{2j+1}(b_a)&=\left[\frac{W_{-j-1,j}W_{-j-2,j+1}}{W_{-j-2,-j-1}W_{j,j+1}}(b_a)\right]^{[+1]},\quad j\in \mathbb{Z}_{\geq 0}.
\ea
\ee
These can be  written as
\beg
{\cal Y}_{s}(b_a)&=\Big[\frac{W_{0,s}W_{-1,s+1}}{W_{-1,0}W_{s,s+1}}(b_a)\Big]^{[-s]},\,\quad s=0,1,2,\ldots.
\label{eq:yfunc2}
\eng
These functions are independent of the normalization of the solution $y_k(z,b_a)$.
Note that ${\cal Y}_0(b_a)=0$ by definition.
By using the Pl\"ucker relation for the determinant of the $2\times2$ matrices:
\begin{align}
W_{k_{1}+1,k_{2}+1}W_{k_{1},k_{2}}=-W_{k_{1}+1,k_{2}}W_{k_{2}+1,k_{1}}-W_{k_{1}+1,k_{1}}W_{k_{2},k_{2}+1},
\end{align}
we find the functions (\ref{eq:yfunc2})  satisfy the relations
\begin{align}\label{eq:Y-system-zb}
{\cal Y}_{s}^{[+1]}(b_a){\cal Y}_{s}^{[-1]}(b_a)&=\Big(1+{\cal Y}_{s-1}(b_a)\Big)\Big(1+{\cal Y}_{s+1}(b_a)\Big), \quad s=0,1,2,\ldots .
\end{align}
This defines the A-type Y-system. 
We then call the function (\ref{eq:yfunc2}) the Y-function.
 Let us consider the Y-function ${\cal Y}_{r+2}$. 
 From the definition (\ref{eq:yfunc2}), $(r+2)$-shift of ${\cal Y}_{r+2}$ is given by
\beg \label{eq:yrplus2}
{\cal Y}_{r+2}^{[r+2]}(b_a)&=\frac{W_{0,r+2}W_{-1,r+3}}{W_{-1,0}W_{r+2,r+3}}(b_a).
\eng
By using the trace formula of monodromy matrix (\ref{eq:trace-mono-Wron}), we find 
\beg
\frac{W_{-1,r+3}}{W_{r+2,r+3}}(b_a)=\omega^{\frac{r+3}{2}}{\rm Tr}[\Omega(b_{a}, l)]^{[-2]}+\frac{W_{0,r+2}}{W_{r+2,r+3}}(b_a).
\eng
Substituting this into (\ref{eq:yrplus2}), we get 
\be
{\cal Y}_{r+2}^{[r+2]}(b_a)
=\omega^{\frac{r+3}{2}}\hat{{\cal Y}}^{[r+2]}{\rm Tr}[\Omega(b_{a}, l)]^{[-2]}+\Big(\hat{{\cal Y}}^{[r+2]}(b_a)\Big)^{2},
\ee
where we have introduced a new Y-function $\hat{{\cal Y}}(b_a)$ by
\be
\hat{{\cal Y}}(b_a)=\Big[\frac{W_{0,r+2}}{W_{-1,0}}(b_a)\Big]^{[-(r+2)]}.
\ee
and used $W_{-1,0}=W_{r+2,r+3}$.
The functional relation for ${\cal Y}_{r+1}$ thus becomes
\be
{\cal Y}_{r+1}^{[+1]}(b_a){\cal Y}_{r+1}^{[-1]}(b_a)
=\Big(1+{\cal Y}_{r}(b_a)\Big)\Big(1+ \omega^{\frac{r+3}{2}} {\rm Tr}[\Omega(b_{a}, l)]^{[-r-4]} \hat{{\cal Y}}(b_a)+\hat{{\cal Y}}(b_a)^{2}\Big).
\ee
Moreover, by using the Pl\"ucker relation we find the relation for $\hat{\cal Y}$.
In summary, we obtain a closed Y-system
\begin{align}
\label{eq:D-type_Y-system}
{\cal Y}_{s}^{[+1]}(b_a) {\cal Y}_{s}^{[-1]}(b_a) &=\Big(1+{\cal Y}_{s-1}(b_a)\Big)\Big(1+{\cal Y}_{s+2}(b_a)\Big),\quad s=1,\cdots,r,\no\\
{\cal Y}_{r+1}^{[+1]}(b_a) {\cal Y}_{r+1}^{[-1]}(b_a) &=\Big(1+{\cal Y}_{r}(b_a) \Big)\Big(1+\omega^{\frac{r+3}{2}}e^{2\pi i l}\hat{{\cal Y}}(b_a) \Big)\Big(1+\omega^{\frac{r+3}{2}}e^{-2\pi i l}\hat{{\cal Y}}(b_a) \Big), 
\\
\hat{{\cal Y}}^{[+1]}(b_a)\hat{{\cal Y}}^{[-1]}(b_a)&=1+{\cal Y}_{r+1}(b_a),\no
\end{align}
where we used the trace of monodromy (\ref{eq:trcae-monodromy}).
Note that this is a $D_{r+3}$ type Y-system. To derive the TBA equations from this Y-system, one needs to consider the functions with different values of $b_a$  at the same time, which is very complicated in general. 
To solve this problem, we will follow the procedure in \cite{Ito:2018eon} where we  introduce the spectral parameter $\zeta$ in the ODE (\ref{eq:Schrodinger-eq-zb}).

\subsection{New Y-system}
Let us introduce the spectral parameter $\zeta$, and rescale the variables in  (\ref{eq:Schrodinger-eq-zb}) as
\begin{align}\label{eq:rescale-su->zb}
x=\zeta^{\frac{2}{r+3}}z,\quad u_{a}&=\zeta^{\frac{2a}{r+3}}b_{a},\quad a=1,2,\cdots, r+2.
\end{align}
In terms of the new variables, the equation (\ref{eq:Schrodinger-eq-zb}) can be written in the form of the Schr{\"o}dinger equation:
\begin{align}\label{eq:Schrodinger-xu}
&\Big(-\zeta^{2}\frac{d^2}{dx^{2}}+x^{r+1}+\sum_{a=1}^{r+2}u_{a}x^{r+1-a}+\frac{\zeta^2 l(l+1)}{x^2}\Big)\hat{\psi}(x,u_a,\zeta)=0,
\end{align}
where the spectral parameter $\zeta$ plays the role of $\hbar$ and the potential is given by
\begin{align}
\label{eq:potential}
V(x)&=Q_0(x)+\zeta^2 Q_2(x), \\
Q_0(x)&=\sum_{a=1}^{r+2}u_{a}x^{r+1-a}, \quad Q_2(x)=\frac{ l(l+1)}{x^2},
\end{align}
where the second term  $\zeta^2 Q_2(x)$ represents the centrifugal potential.
We can then regard the solution $y(z,b_a)$ as the function of $\zeta$, $x$ and $u_a$:
\be
\hat{y}(x,u_{a},\zeta)=y(z,b_{a})=y(\zeta^{-\frac{2}{r+2}}x,-\zeta^{-\frac{2(a+1)}{r+3}}u_{a}), \quad a=1,\cdots,r+2.
\ee
The action of the Symanzik rotation $(z,b_a, l)\to (\omega^{-k} z,\omega^{-ka}b_a, l)$ can be expressed by the shift of the phase of $\zeta$:
\begin{align}
(\omega^{-k}z,\omega^{-ak}b_{a}, l)=\left((e^{i\pi k}\zeta)^{-\frac{2}{r+3}}x,(e^{i\pi k}\zeta)^{-\frac{2a}{r+3}}u_{a}, l)\right),\quad a=1,\cdots,r+2.
\end{align}
The solutions $y_k(z,b_a)$ thus can be obtained from $\hat{y}(x, u_a,\zeta)$ by using
\be
\hat{y}_{k}(x,u_a,\zeta)=\omega^{\frac{k}{2}}\hat{y}(x,u_{a},e^{i\pi k}\zeta).
\ee
We also introduce the Wronskian of $\hat{y}_j$ by
\begin{align}
\hat{W}_{k_1,k_2}(\zeta,u_a)=\zeta^{\frac{2}{r+3}}\Big(\hat{y}_{k_1}(x,u_a,\zeta) \partial_x\hat{y}_{k_2}(x,u_a,\zeta)-\hat{y}_{k_2}(x,u_a,\zeta) \partial_x\hat{y}_{k_1}(x,u_a,\zeta)\Big).
\end{align}
This  is normalized such that it agrees with $W_{k_1,k_2}(b_a)$. 
Using the Wronskians, we redefine the Y-functions ${\cal Y}_s(b_a)$ and $\hat{\cal Y}(b_a)$ as
the functions of $\zeta$ and $u_a$, which are denoted as $Y_s(\zeta,u_a)$ and $\hat{Y}(\zeta,u_a)$,
respectively.
Explicitly they are defined by
\begin{align}
Y_{2j}(\zeta,u_a)&=\frac{\hat{W}_{-j,j}\hat{W}_{-j-1,j+1}}{\hat{W}_{-j-1,-j}\hat{W}_{j,j+1}}(\zeta,u_a),\no\\
Y_{2j+1}(e^{-\frac{\pi i}{2}}\zeta,u_a)&=\frac{\hat{W}_{-j-1,j}\hat{W}_{-j-2,j+1}}{\hat{W}_{-j-2,-j-1}\hat{W}_{j,j+1}}(\zeta,u_a),\label{eq:new-Y}\\
\hat{Y}(\zeta,u_{a})&=\frac{\hat{W}_{0,r+2}}{\hat{W}_{-1,0}}(e^{-\frac{i\pi}{2}(r+2)}\zeta, u_a)\no.
\end{align}
Now the Y-system (\ref{eq:D-type_Y-system}) can be written as
\be
\ba
\label{eq:Y-system-xu}
Y_{s}(e^{\frac{\pi i}{2}}\zeta,u_{a})Y_{s}(e^{-\frac{\pi i}{2}}\zeta,u_{a})&=\Big(1+Y_{s-1}(\zeta,u_{a})\Big)\Big(1+Y_{s+2}(\zeta,u_{a})\Big),\quad s=1,\cdots,r,\\
Y_{r+1}(e^{\frac{\pi i}{2}}\zeta,u_{a})Y_{r+1}(e^{-\frac{\pi i}{2}}\zeta,u_{a})&=\Big(1+Y_{r}(\zeta,u_{a})\Big)\Big(1+\omega^{\frac{r+3}{2}} e^{2\pi i l}\hat{Y}(\zeta,u_{a})\Big)\Big(1+\omega^{\frac{r+3}{2}} e^{-2\pi i l}\hat{Y}(\zeta,u_{a})\Big),\\
\hat{Y}(e^{\frac{\pi i}{2}}\zeta,u_{a})\hat{Y}(e^{-\frac{\pi i}{2}}\zeta,u_{a})&=1+Y_{r+1}(\zeta,u_{a}),
\ea
\ee 
where $Y_0(\zeta,u_a)=0$. Since in this Y-system the Y-functions depend on the same $u_a$'s, we will omit the argument $u_a$ in the Y-functions. 

The Y-system (\ref{eq:Y-system-zb}) and (\ref{eq:Y-system-xu}) are essentially the same system, but interpretations of the spectral parameter are different. The Y-system (\ref{eq:Y-system-xu}) also appears in the study of the integrable structure in the minimal model \cite{Bazhanov:1996aq}, in the massive ODE/IM correspondence for the modified sinh-Gordon equation \cite{Lukyanov:2010rn} and in the study of the minimal surface in AdS$_3$ corresponding to the form factor \cite{Maldacena:2010kp}.\footnote{See also \cite{Gao:2013dza, Ito:2016qzt} for examples.}


\section{TBA system}
From the analytical properties and asymptotic behaviors of the Y-functions, one can derive a set of non-linear integral equations satisfied by the Y-functions, which is called the Thermodynamic Bethe Ansatz (TBA) equations \cite{Yang:1968rm,Zamolodchikov:1989cf}. 
Our TBA system turns out to be a massless version of these TBA equations. 
Similar massive TBA equations can be obtained from the Hitchin system in the study of AdS${}_3$ minimal surface and form factors  \cite{Alday:2010vh, Maldacena:2010kp}.

\subsection{TBA equations}\label{sec:TBA}

To derive the TBA equations from the Y-system (\ref{eq:Y-system-xu}), we first need to investigate the small $\zeta$ behavior of the Y-functions. The WKB expansion  of $\hat{y}_k(x,u_a,\zeta)$ is given by 
\be
\label{eq:y-exact-WKB}
\hat{y}_k(x,u_a,\zeta)=(-1)^{\frac{k}{2}}c(\zeta)\exp\left(\frac{{\delta}_k}{\zeta}\int^{x}_{x_k}P(x^\prime)dx^\prime\right),
\ee
where $c(\zeta)=\frac{1}{\sqrt{2i}}\zeta^{\frac{(r+1)}{2(r+3)}}$, and $P(x)$ is expanded in $\zeta$ as $P(x,\zeta)=\sum_{n=0}^\infty\zeta^n P_n(x)$. 
From the differential equation (\ref{eq:Schrodinger-xu}), $P(x,\zeta)$ satisfies
\begin{align}\label{eq:WKB-recursive}
P(x,\zeta)^2+\delta_k \zeta P'(x,\zeta)-V(x)=0.
\end{align}
We find that  $P_0(x)=\sqrt{Q_0(x)}$ and 
 $P_n(x)$ $(n\geq 1)$ can be determined recursively. $P_{odd}(x,\zeta)=\sum_{n=0}^{\infty}\zeta^{2n+1}P_{2n+1}(x)$ is determined by $P_{even}(x,\zeta)=\sum_{n=0}^{\infty}\zeta^{2n}P_{2n}(x)$ as
 $P_{odd}(x)=-\frac{\delta_k}{2}(\log P_{even}(x,\zeta))'$. 
In the exponential of (\ref{eq:y-exact-WKB}), ${\delta}_k=\pm (-1)^k$ denotes the sign factor, where $\pm$ depends on the plus (minus) sheet of the Riemann surface (the WKB curve)
defined by
\begin{align}
    y^2=Q_0(x)=x^{r+1}+\cdots+\frac{u_{r+2}}{x}.
    \label{eq:wkbcurve}
\end{align}
We evaluate the Wronskian $\hat{W}_{k_1,k_2}(\zeta)$ by using the WKB solution (\ref{eq:y-exact-WKB}):
\begin{equation*}\label{eq:Wki}
\hat{W}_{k,j}=i(-1)^{\frac{k+j}{2}}\delta_k\exp\left(\frac{\delta_k}{\zeta}\int^{x_j}_{x_k}P_{even}(x^\prime)d x^\prime+\frac{1}{2}[\log P_{even}(x_j)+\log P_{even}(x_k)]\right).
\end{equation*}
By choosing the initial points of the solution, we can express the Y-function (\ref{eq:new-Y})  by using a contour integral over a certain 1-cycle on the curve (\ref{eq:wkbcurve}). We find that the Y-functions at small $\zeta$ behave as
\be\label{eq:logyasymp}
\ba
\log{Y}_{2k+1}(\zeta)&\sim -\frac{1}{i\zeta}\oint_{\gamma_{2k+1}}P_0(x)dx=:-\frac{m_{2k+1}}{\zeta},\\
 \log{Y}_{2k}(\zeta)&\sim -\frac{1}{\zeta}\oint_{\gamma_{2k}}P_0(x)dx=:-\frac{m_{2k},}{\zeta},\\
\log\hat{Y}(\zeta)&\sim \begin{cases}
-\frac{1}{\zeta}\oint_{\hat{\gamma}}P_{0}(x)dx=:-\frac{\hat{m}}{\zeta} & r:\mbox{even}\\
-\frac{1}{i\zeta}\oint_{\hat{\gamma}}P_{0}(x)dx=:-\frac{\hat{m}}{\zeta} & r:\mbox{odd}
\end{cases}.
\ea
\ee
Note that the above approximation is valid in the sector $|\mbox{arg}(\zeta)|<\pi$. The classical periods $m_s$ are functions of moduli parameters $(u_1,\cdots,u_{r+2})$. There are $r+2$ independent period integrals for $m_s$ for generic $u_a$'s. 
$\gamma_s$ and $\hat{\gamma}$ are the one-cycles on the Riemann surface (\ref{eq:wkbcurve}), 
which possess $r+2$ turning points and one simple pole. For simplicity, we will first consider the case such that  all the masses $m_s$ are positive and real.

\begin{figure}[tb]
\begin{center}
\begin{tabular}{cc}
\resizebox{60mm}{!}{\includegraphics{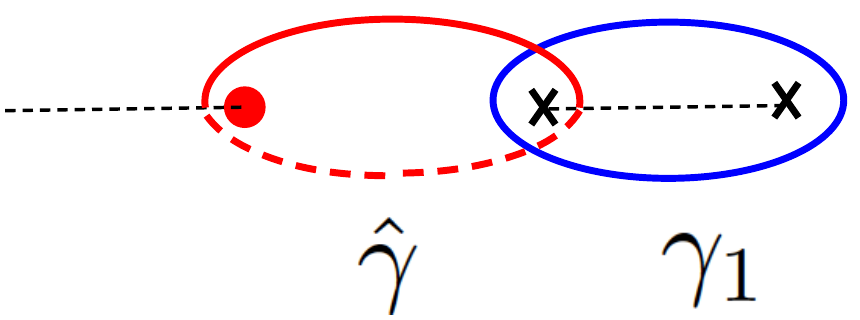}}
\hspace{10mm}
&
\resizebox{60mm}{!}{\includegraphics{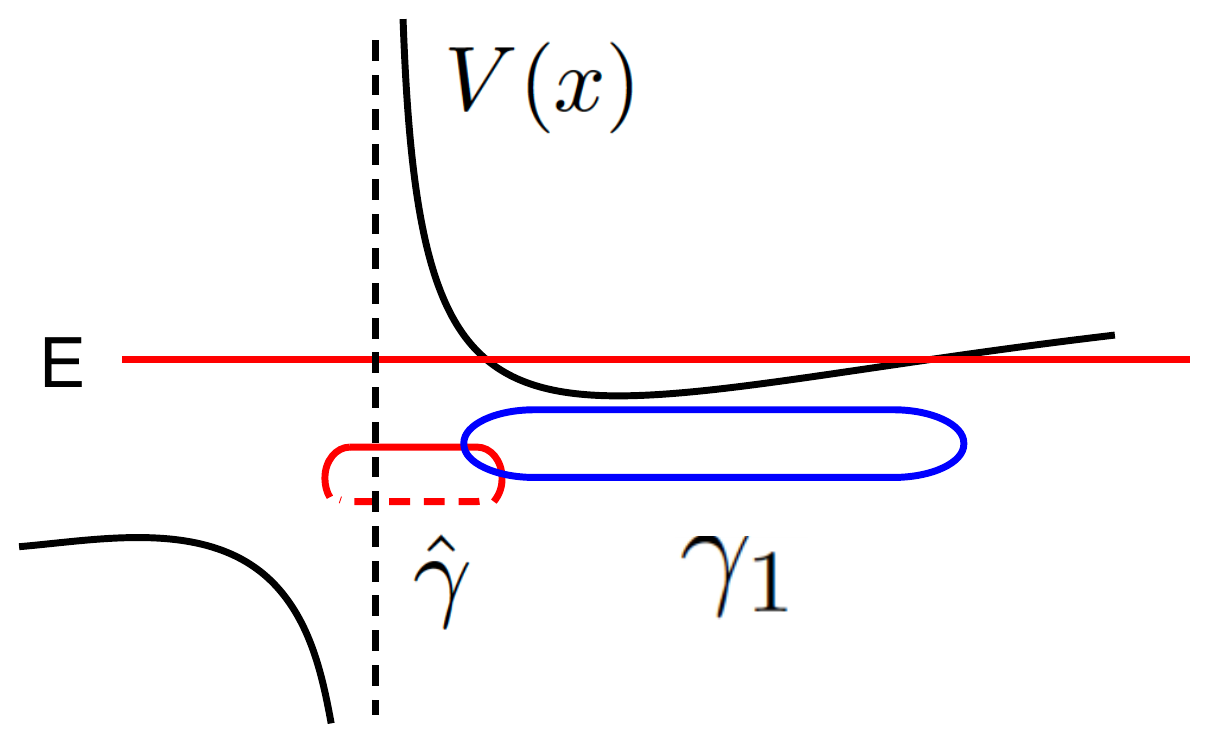}}
\end{tabular}
\end{center}
  \caption{The contour of the Y-functions ($r=0$) on the Riemann surface $y^2=\frac{x^2+u_1x+u_{2}}{x}$ is shown on the left. The crosses indicate the turning points satisfying $x^2+u_1x+u_{2}=0$. The red dot is the simple pole $x=0$. The dotted line is the branch cut on the Riemann surface. 
  On the right, we show the corresponding potential and the cycles, where $\gamma_1$ and $\hat{\gamma}$ are the classically allowed cycle and classically forbidden cycle respectively.}
\label{fig:r0}
\end{figure}

\begin{figure}[tb]
\begin{center}
\begin{tabular}{cc}
\resizebox{60mm}{!}{\includegraphics{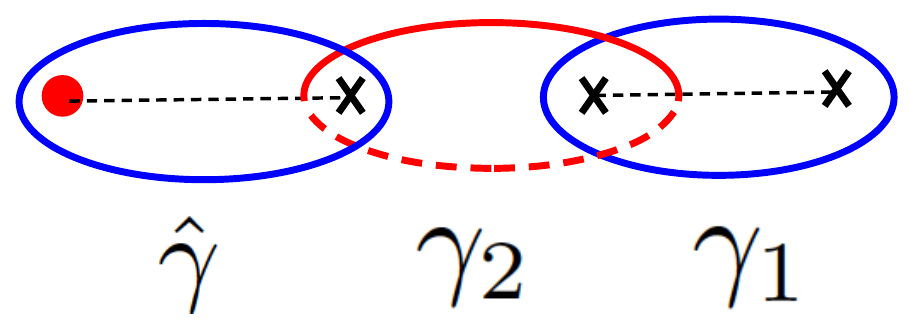}}
\hspace{10mm}
&
\resizebox{70mm}{!}{\includegraphics{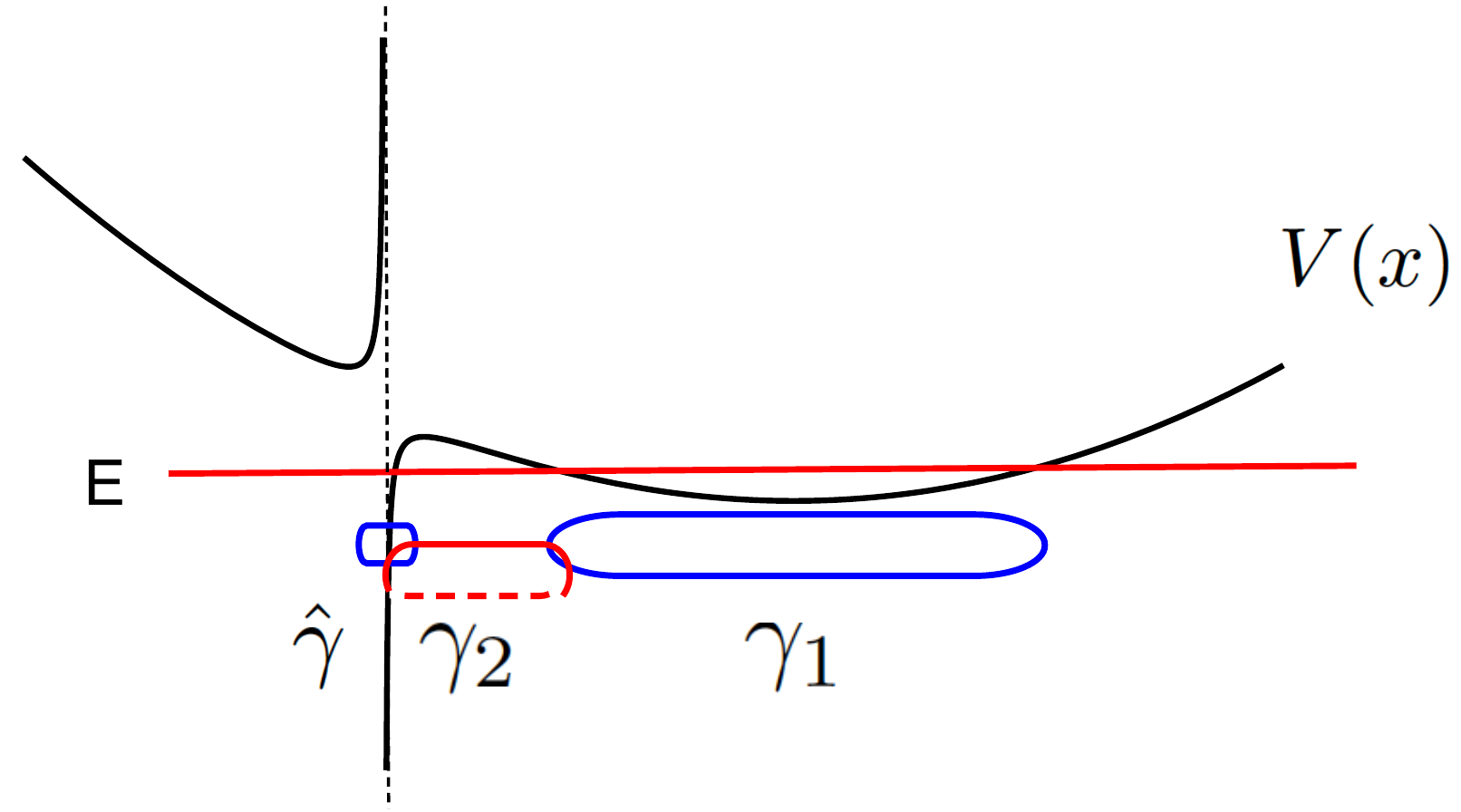}}
\end{tabular}
\end{center}
  \caption{The contours of the Y-function ($r=1$) on the Riemann surface $y^2=\frac{x^3+u_1x^2+u_{2}x+u_3}{x}$ are shown on the left. The crosses indicate the turning points satisfying $x^3+u_1x^2+u_{2}x+u_3=0$. The red dot is the simple pole $x=0$. The dotted line is the branch cut on the Riemann surface. 
  On the right, we show the corresponding potential and the cycles, $\gamma_1$ and $\hat{\gamma}$ are the classically allowed cycles, and $\gamma_2$ classically forbidden cycle.}
\label{fig:r1}
\end{figure}

We choose one-cycles $\gamma_s$ and $\hat{\gamma}$ on the curve (\ref{eq:wkbcurve}) such that the intersection number of $\gamma_s$ and $\hat{\gamma}$ as $\langle \gamma_{s'},\gamma_s\rangle=\delta_{s',s+1}-\delta_{s',s-1}$ and 
$\langle \hat{\gamma}, \gamma_{r+1}\rangle=1$.
We show examples of the contours for the $r=0$ and $r=1$ case in Fig. \ref{fig:r0} and Fig. \ref{fig:r1}. 
For $r=0$, the crossed points indicate the roots of $x^2+u_1 x+u_2$, which are assumed to be real and positive. The cycle $\gamma_1$ encircle the cut between these turning points. While the cycle $\hat{\gamma}$ encircles the simple pole $x=0$ and
one of the root. 
For $r=1$  the contours are given as in the  $r=0$ case.

\begin{figure}[tbh]
\begin{center}
\resizebox{55mm}{!}{\includegraphics{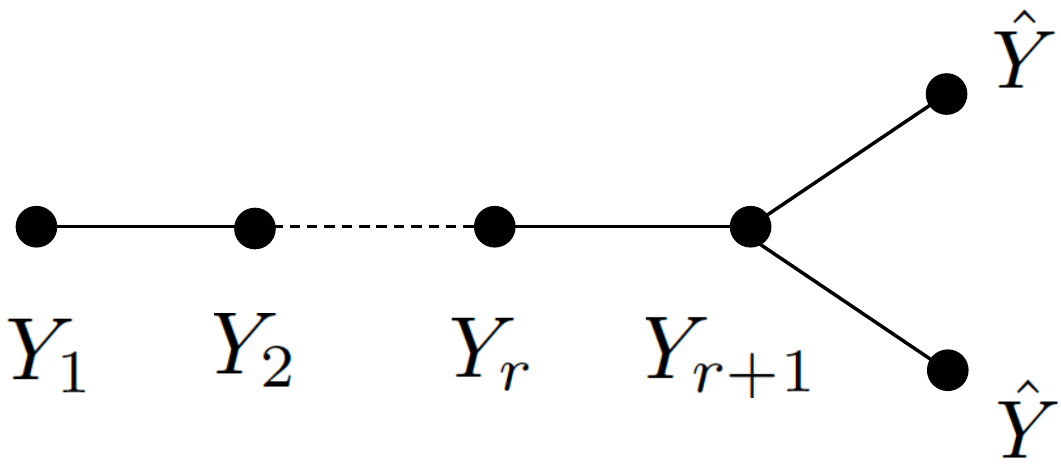}}
\end{center}
  \caption{The Dynkin diagram for $D_{r+3}$-type TBA system.}
\label{fig:Dyn-dym}
\end{figure}

We then introduce the  functions $\ell_s(\zeta)$ and $\hat{\ell}(\zeta)$ of $\zeta$, which is analytic in $|\arg(\zeta)|\leq\frac{\pi}{2}$:
\be
\ell_s(\zeta):=\log Y_s(\zeta)+\frac{m_s}{\zeta},\quad \hat{\ell}(\zeta)=\log\hat{Y}(\zeta)+\frac{\hat{m}}{\zeta}.
\ee
We can check the analyticity of the Y-function numerically as we will see in the following sections. 
The Y-system then leads to
\be
\ba
\label{eq:Y-system-ell}
\ell_{s}(e^{\frac{\pi i}{2}}\zeta)+\ell_{s}(e^{-\frac{\pi i}{2}}\zeta)&=\log\Big(\big(1+Y_{s+1}(\zeta)\big)\big(1+Y_{s-1}(\zeta)\big)\Big),\quad s=1,\cdots,r,\\
\ell_{r+1}(e^{\frac{\pi i}{2}}\zeta)+\ell_{r+1}(e^{-\frac{\pi i}{2}}\zeta)&=\log\Big(\big(1+Y_{r}(\zeta)\big)\big(1+\omega^{\frac{r+3}{2}} e^{2\pi i l}\hat{Y}(\zeta)\big)\big(1+\omega^{\frac{r+3}{2}} e^{-2\pi i l}\hat{Y}(\zeta)\big)\Big),\\
\hat{\ell}(e^{\frac{\pi i}{2}}\zeta)+\hat{\ell}(e^{-\frac{\pi i}{2}}\zeta)&=\log\Big(1+Y_{r+1}(\zeta,u_{a})\Big).
\ea
\ee
Here we set $\zeta=e^{-\theta}$ and introduce the kernel
\be
K(\theta)=\frac{1}{2\pi \cosh(\theta)}.
\ee
Convoluting (\ref{eq:Y-system-ell}) with the kernel $K(\theta'-\theta)$ on the real axis, the l.h.s picks a pole at $\theta'=\theta$.  One derives the TBA equations
\be
\ba
\label{eq:TBA-pole}
\log Y_{s}(\theta)&=-m_{s}e^{\theta}+\int_{\mathbb{R}}\frac{\log\Big(1+Y_{s-1}(\theta^{\prime})\Big)+\log\Big(1+Y_{s+1}(\theta^{\prime})\Big) }{\cosh(\theta-\theta^{\prime})}\frac{{d}\theta^{\prime}}{2\pi},\quad s=1,\cdots,r,\\
\log Y_{r+1}(\theta)&=-m_{r+1}e^{\theta}+\int_{\mathbb{R}}\frac{\log\Big(1+Y_{r}(\theta^{\prime})\Big)}{\cosh(\theta-\theta^{\prime})}\frac{{d}\theta^{\prime}}{2\pi}
+\int_{\mathbb{R}}\frac{\log\Big(\big(1+\omega^{\frac{r+3}{2}}e^{2\pi i l}\hat{Y}(\theta^{\prime})\big) \big(1+\omega^{\frac{r+3}{2}}e^{-2\pi i l}\hat{Y}(\theta^{\prime})\big)\Big)}{\cosh(\theta-\theta^{\prime})}\frac{{d}\theta^{\prime}}{2\pi},\\
\log\hat{Y}(\theta)&=-\hat{m}e^{\theta}+\int_{\mathbb{R}}\frac{\log\Big(1+Y_{r+1}(\theta^{\prime})\Big)}{\cosh(\theta-\theta^{\prime})}\frac{{d}\theta^{\prime}}{2\pi}.
\ea
\ee
This TBA system corresponds to the $D_{r+3}$-type Dynkin diagram Fig.\ref{fig:Dyn-dym}.

The TBA system (\ref{eq:TBA-pole}) can be regarded as the ``conformal limit'' of the TBA system considered in \cite{Maldacena:2010kp} but with a specific monodromy. Note that in the above derivation, we have assumed that all the masses $m_s$ and $\hat{m}$ to be positive and real. We will explain the TBA equations with complex masses at the end of this section.

\subsection{Large $\theta$ behavior of Y-function}
At large $\theta$, the TBA equations (\ref{eq:TBA-pole}) lead to the 
 expansion of Y-functions:
\be
\ba
\label{eq:Y-exp}
-\log Y_{s}(\theta)&=m_{s}e^{\theta}+\sum_{n\geq1}m_{s}^{(n)}e^{(1-2n)\theta},\quad s=1,\cdots,r+1,\\
-\log\hat{Y}(\theta)&=\hat{m}e^{\theta}+\sum_{n\geq1}\hat{m}^{(n)}e^{(1-2n)\theta},
\ea
\ee
where
\be
\ba
\label{eq:mn}
m_{s}^{(n)}&=\frac{(-1)^{n}}{\pi}\int_{\mathbb{R}}e^{(2n-1)\theta}\Big\{\log\Big(1+Y_{s-1}(\theta^{\prime})\Big)+\log\Big(1+Y_{s+1}(\theta^{\prime})\Big)\Big\}{d}\theta^{\prime},\quad s=1,\cdots,r,\\
m_{r+1}^{(n)}&=\frac{(-1)^{n}}{\pi}\int_{\mathbb{R}}e^{(2n-1)\theta}\Big\{\log\Big(1+Y_{r}(\theta^{\prime})\Big)+\log\Big(1+\omega^{\frac{r+3}{2}}e^{2\pi il}\hat{Y}(\theta^{\prime})\Big)+\log\Big(1+\omega^{\frac{r+3}{2}}e^{-2\pi il}\hat{Y}(\theta^{\prime})\Big)\Big\}{d}\theta^{\prime},\\
\hat{m}^{(n)}&=\frac{(-1)^{n}}{\pi}\int_{\mathbb{R}}e^{(2n-1)\theta}\log\Big(1+Y_{r+1}(\theta^{\prime})\Big){d}\theta^{\prime}.
\ea
\ee
By solving the TBA equations (\ref{eq:TBA-pole})  numerically, one obtains $m_s^{(n)}$ ($s=1,\cdots,r+1$) 
and $\hat{m}^{(n)}$. In this way, we can extract the perturbative series of Y-function from the TBA equations. In the next section, we will compare the expansion of Y-functions with the expansion of the WKB periods.

\subsection{Effective central charge and ``massless'' TBA}
We can also take the UV limit $\theta\rightarrow -\infty$.
As $\theta\to -\infty$, the Y-function approaches to a constant value
$Y_s\rightarrow Y^*_s$, $\hat{Y}\rightarrow \hat{Y}^*$, where
\be
\ba
Y_{s}^\ast=\frac{\sin(\frac{\pi(2l+1)}{(r+3)}(s+2))\sin(\frac{\pi(2l+1)}{(r+3)}s))}{\sin^{2}(\frac{\pi(2l+1)}{(r+3)})},\quad \hat{Y}^\ast=\frac{\sin(\frac{\pi(2l+1)(r+2)}{(r+3)})}{\sin(\frac{\pi(2l+1)}{(r+3)})}.
\ea
\ee
The  ``effective central charge'' associated to our TBA system is defined by  
\be
\ba
c_{{\rm eff}}&=\frac{6}{\pi^{2}}\sum_{s=1}^{r+1}m_{s}\int e^{\theta}\log\Big(1+Y_{s}(\theta)\Big)d\theta+\frac{6}{\pi^{2}}\hat{m}\int e^{\theta}\log\Big(\big(1+\omega^{\frac{r+3}{2}e^{2\pi i l}}\hat{Y}(\theta)\big) \big(1+\omega^{\frac{r+3}{2}e^{-2\pi i l}}\hat{Y}(\theta)\big) \Big). 
\ea\ee
In the UV limit, it can be evaluated as 
\be\ba
c_{\rm eff}&=\frac{6}{\pi^{2}}\Big(\sum_{s=1}^{r+1}{\cal L}_{1}(\frac{1}{1+\frac{1}{Y_{s}^{\ast}}})+{\cal L}_{-e^{2\pi il}}(\frac{1}{-e^{2\pi il}+\frac{1}{\hat{Y}^{\ast}}})+{\cal L}_{-e^{-2\pi il}}(\frac{1}{-e^{-2\pi il}+\frac{1}{\hat{Y}^{\ast}}})\Big)
\label{eq:ceff-diag},
\ea
\ee
where 
\be
\ba
{\cal L}_{c}(x)&=-\frac{1}{2}\int_{0}^{x}dy\Big(\frac{c\log y}{1-cy}+\frac{\log(1-cy)}{y}\Big)=\frac{1}{2}\Big(\log x\log(1-cx)+2{\rm Li}_{2}(cx)\Big).
\ea
\ee
and ${\rm Li}_2(x)$ is the dilogarithm function. From the dilogarithm identities \cite{Kirillov:1987} we find the effective central charge is
\be
\label{eq:ceff}
c_{\rm eff}(r)=(r+2)\Big(1-24\frac{(l+\frac{1}{2})^2}{r+3}\Big).
\ee

\begin{figure}[tbh]
\begin{center}
\resizebox{55mm}{!}{\includegraphics{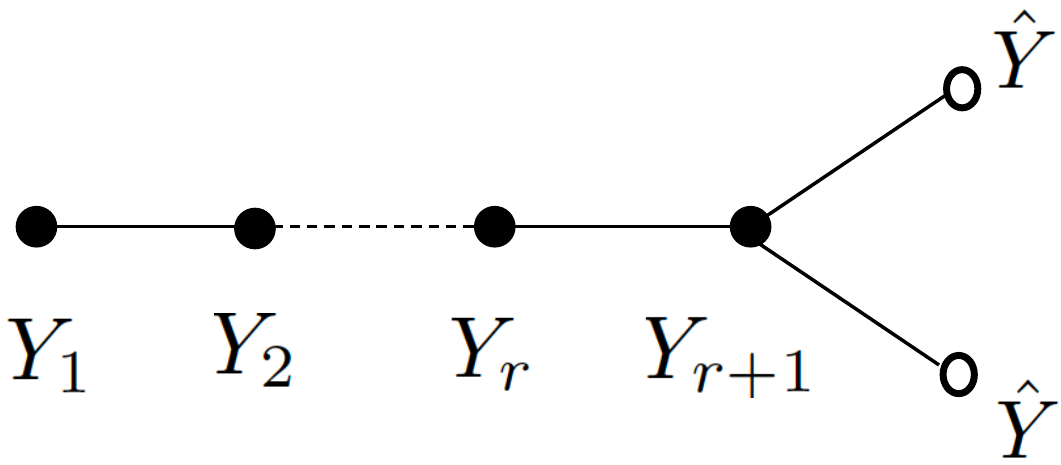}}
\end{center}
  \caption{The Dynkin diagram for $D_{r+3}$-type TBA system with massless $\hat{Y}$.}
\label{fig:Dyn-dym-massless}
\end{figure}

It is interesting to consider the limit $u_{r+1}\rightarrow 0$ of the TBA equations, where the particle corresponding to 
 $\hat{Y}$ becomes massless, {\it i.e.} $\hat{m}=0$.
 We may label this TBA equations by using the Dynkin diagram in Fig.\ref{fig:Dyn-dym-massless}.
In this case, the constant solution of $\hat{Y}$ will also contribute to the UV limit of the effective central charge of the integrable model. Subtracting this contribution, we thus obtain the effective central charge
\be
\ba 
c_{\rm eff}(r)\big|_{u_{r+2}=0}&=c_{{\rm eff},D_{r+3}}-c_{{\rm eff},D_{2}}=c_{{\rm eff}}(r)-c_{{\rm eff}}(-1)\\&=r+1-12\frac{r+1}{r+3}(l+\frac{1}{2})^{2}.
\ea
\ee
Taking the limit $l\to 0$, we obtain
\be
\label{eq:ceff-massless}
\lim_{l\to0}c_{\rm eff}(r)\big|_{u_{r+2}=0}=r+1-3\frac{r+1}{r+3}=\frac{r(r+1)}{r+3},
\ee
which coincides with the one obtained in \cite{Ito:2018eon}. When $r=0$ and $l\neq 0$, the Schr\"odinger equation (\ref{eq:Schrodinger-eq-zb}) recovers the one studied in \cite{Bazhanov:1998wj, Dorey:1999uk} and \cite{Dorey:2007zx} for the potential $z+\frac{l(l+1)}{z^2}$. In this case the potential is
\be
\lim_{l\to0}c_{\rm eff}(r=0)\big|_{u_{r+2}=0}=1-4(l+\frac{1}{2})^2,
\ee
which reproduces the effective central charge (D.60) in \cite{Dorey:2007zx} with $M=1/2$.
We note that 
when the TBA system has extra discrete symmetry at some points in the moduli space, the central charge must be divided by the discrete symmetry.

\subsection{TBA equations for complex mass in minimal chamber}
\begin{figure}[tbh]
\begin{center}
\resizebox{40mm}{!}{\includegraphics{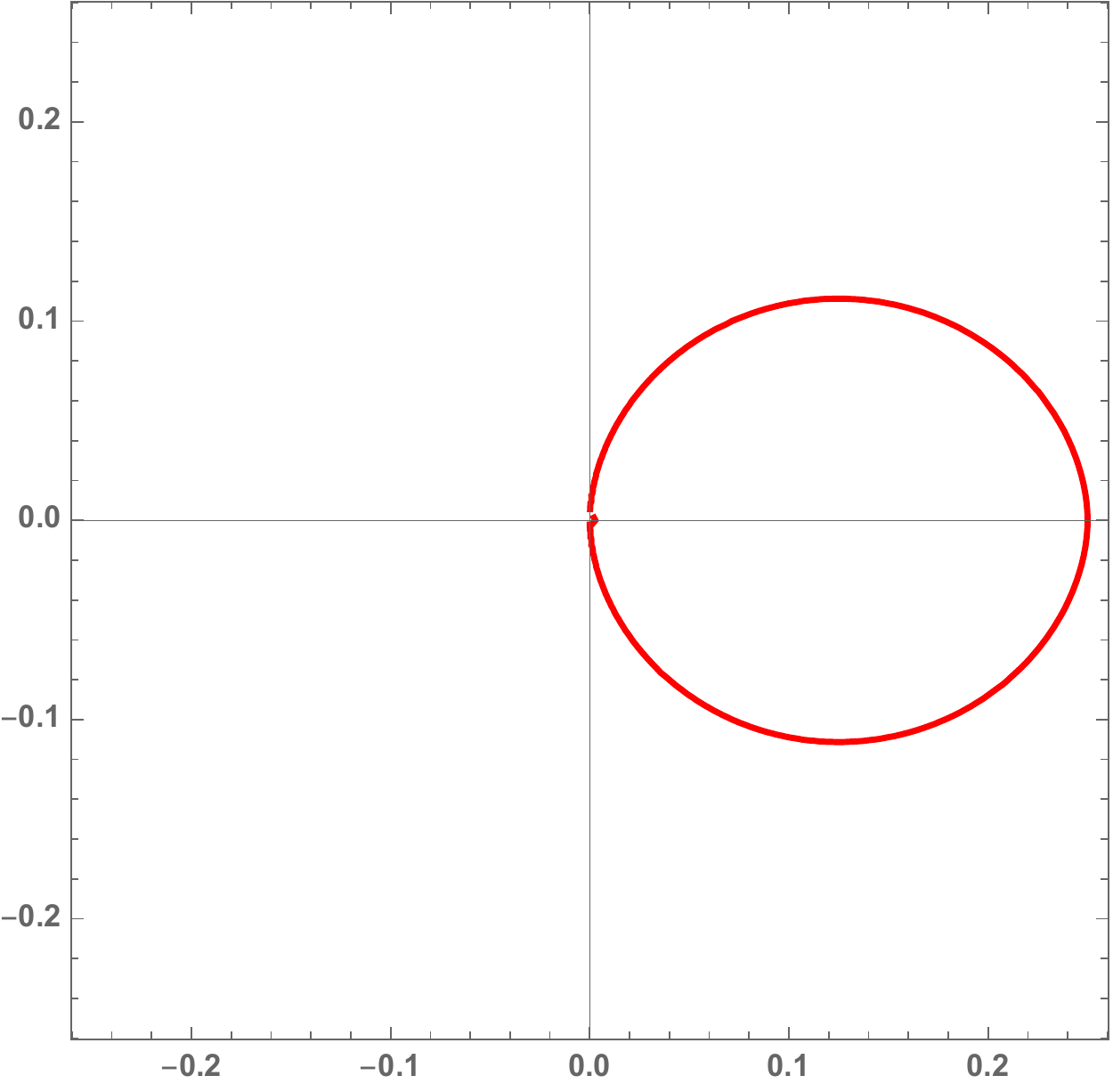}}
\end{center}
  \caption{The wall of marginal stability for the cade $Q_0(x)=\frac{x^2-x+u_2}{x}$, in the complex $u_2$ plane.}
\label{fig:wall-u2}
\end{figure}

The derivation we presented in section \ref{sec:TBA} is valid for the case with positive and real masses. We now extend the TBA equations (\ref{eq:TBA-pole}) to complex masses by following the procedure in \cite{Alday:2010vh}. Let us denote the masses by
\be
m_{s}=|m_s|e^{i\phi_s},\quad \hat{m}=|\hat{m}|e^{i \hat{\phi}},\quad s=1,\cdots, r+1.
\ee
We then shift the argument of the Y-function, such that the masses of the shifted Y-functions $Y_s(\theta-i\phi_1
)$ and $\hat{Y}(\theta-i\hat{\phi}
)$ are positive and real. Therefore we obtain the TBA equations:
\be
\ba
\label{eq:TBA-com-mass}
\log Y_{s}(\theta-i\phi_{1})&=-|m_{s}|e^{\theta}+\int_{\mathbb{R}}\frac{\log\Big(1+Y_{s-1}(\theta^{\prime}-i\phi_{s-1})\Big)}{\cosh(\theta-\theta^{\prime}-i\phi_{s}+i\phi_{s-1})}\frac{{d}\theta^{\prime}}{2\pi}\\
&\qquad+\int_{\mathbb{R}}\frac{\log\Big(1+Y_{s+1}(\theta^{\prime}-i\phi_{s+1})\Big)}{\cosh(\theta-\theta^{\prime}-i\phi_{s}+i\phi_{s+1})}\frac{{d}\theta^{\prime}}{2\pi},\quad s=1,\cdots,r,\\
\log Y_{r+1}(\theta-i\phi_{r+1})&=-|m_{r+1}|e^{\theta}+\int_{\mathbb{R}}\frac{\log\Big(1+Y_{r}(\theta^{\prime}-i\phi_{r})\Big)}{\cosh(\theta-\theta^{\prime}-i\phi_{r+1}+i\phi_{r})}\frac{{d}\theta^{\prime}}{2\pi}\\
&\qquad+\int_{\mathbb{R}}\frac{\log\Big(1+\omega^{\frac{r+3}{2}}e^{2\pi il}\hat{Y}(\theta^{\prime}-i\hat{\phi})\Big)}{\cosh(\theta-\theta^{\prime}-i\phi_{r+1}+i\hat{\phi})}\frac{{d}\theta^{\prime}}{2\pi}+\int_{\mathbb{R}}\frac{\log\Big(1+\omega^{\frac{r+3}{2}}e^{-2\pi il}\hat{Y}(\theta^{\prime}-i\hat{\phi})\Big)}{\cosh(\theta-\theta^{\prime}-i\phi_{r+1}+i\hat{\phi})}\frac{{d}\theta^{\prime}}{2\pi},\\
\log\hat{Y}(\theta-i\hat{\phi})&=-|\hat{m}|e^{\theta}+\int_{\mathbb{R}}\frac{\log\Big(1+Y_{r+1}(\theta^{\prime}-i\phi_{r+1})\Big)}{\cosh(\theta-\theta^{\prime}-i\hat{\phi}+i\phi_{r+1})}\frac{{d}\theta^{\prime}}{2\pi}.
\ea
\ee
Note that poles appear in the kernel of these TBA equations when the phases satisfy the conditions:
\be
\label{eq:pole-kernel}
|\phi_s-\phi_{s\pm 1}|=\frac{\pi}{2}\quad {\rm or}\quad |\phi_{r+1}-\hat{\phi}|=\frac{\pi}{2}.
\ee
Hence the TBA equations (\ref{eq:TBA-com-mass}) are  valid in the region
\be
\label{eq:minimal-chamber}
|\phi_s-\phi_{s\pm 1}|<\frac{\pi}{2}\quad {\rm and}\quad |\phi_{r+1}-\hat{\phi}|<\frac{\pi}{2}.
\ee
This region is called the ``minimal chamber''  in the moduli space, whose boundary defines the marginal stability wall \cite{Gaiotto:2008cd, Gaiotto:2009hg}. 
In Fig. \ref{fig:wall-u2}, we plot the curve  of marginal stability for the case $Q_0(x)=\frac{x^2-x+u_2}{x}$, in the complex $u_2$ plane. The region inside of the wall is the ``minimal chamber'' for $r=0$ case. 

As $\phi_s-\phi_{s\pm 1}$ or $\phi_{r+1}-\hat{\phi}$ take the values $\pm\frac{\pi}{2}$, $\pm\frac{3\pi}{2}$, we should modify the form of the TBA equations to include the contribution of the poles (\ref{eq:pole-kernel}) in the kernel, where we need to consider the wall-crossing of the TBA equations. More details on the wall crossing will be presented in the appendix \ref{app:WC}.


\section{TBA equations and WKB periods}
In the previous work\cite{Ito:2018eon}, we  checked numerically that the WKB periods agree with the the large $\theta$-expansion of the logarithm of the Y-function for polynomial potential. In this subsection we confirm this relation for the potential with a regular singularity. The WKB periods $\Pi_\gamma(\zeta)$ for the Schr\"odinger equation (\ref{eq:Schrodinger-xu}) are defined by the period integral of $P(x,\zeta)$, which are formal even power series of $\zeta$:
\be
\label{eq:WKB-series}
\Pi_\gamma(\zeta)=\sum_{n\geq 0}\Pi_\gamma^{(n)}\zeta^{2n},\quad \gamma=\gamma_s,\hat{\gamma},\quad s=1,\cdots,r+1.
\ee
The classical periods $\Pi^{(0)}_{\gamma_s}$ and $\Pi^{(0)}_{\hat{\gamma}}$ are determined  $m_s$ and $\hat{m}$ respectively using the formulas (\ref{eq:logyasymp}). The $\zeta^2$ correction $\Pi^{(1)}_\gamma$ and the $\zeta^4$ correction $\Pi^{(2)}_\gamma$ of the WKB period are given by
\be
\label{eq:Pi1}
\Pi^{(1)}_\gamma=\oint_\gamma dx\Big(\frac{1}{2}\frac{Q_{2}}{\sqrt{Q_{0}}}+\frac{1}{48}\frac{\partial_{x}^{2}Q_0}{Q_0(x)^{3/2}}\Big)
\ee
and
\be
\ba
\label{eq:Pi2}
\Pi^{(2)}_\gamma=\oint_\gamma dx\Big(-\frac{7}{1536}\frac{(\partial_{x}^{2}Q_{0})^{2}}{Q_{0}^{7/2}}+\frac{1}{768}\frac{\partial_{x}^{4}Q_{0}}{Q_{0}^{5/2}}-\frac{Q_{2}\partial_{x}^{2}Q_{0}}{32Q_{0}^{5/2}}+\frac{\partial_{x}^{2}Q_{2}}{48Q_{0}^{3/2}}-\frac{Q_{2}^{2}}{8Q_{0}^{3/2}}\Big)
\ea
\ee
respectively \cite{Ito:2017iba}. These higher-order corrections to the WKB periods can be obtained by acting the differential operator on the classical periods.
In the present case, 
following \cite{Ito:2017iba, Ito:2018hwp, Ito:2019twh}, we find the  formulas
\begin{align}
\label{eq:Pi1-exp}
\Pi^{(1)}_\gamma
&=-{1\over12} \sum_{k=0}^{r+1} u_k (r+2-k)(r-k-3) \partial_{u_{k+1}}\partial_{u_{r+2}}\Pi^{(0)}_\gamma
+{l(l+1)\over u_{r+2}} \sum_{k=0}^{r+1} u_k (r+1-k) \partial_{u_{k+1}} 
\Pi^{(0)}_{\gamma},
\\
\Pi^{(2)}_{\gamma}&={7\over 1440}
\Bigl[
\sum_{i,j=0}^{r-1
}  [r+1-i]_2 [r+1-j]_2 
u_i u_j \partial_{u_{i+1
}} \partial_{u_{j+1
}}
\partial_{u_{r+2}}^2
\nonumber\\
&+4 u_{r+2} 
\sum_{i=0}^{r
} u_i \left\{ [r+1-i]_2 
-{1\over3} (5r-5i+13) 
\right\} \partial_{u_{i+2}}\partial_{u_{r+2}}^3
\nonumber\\
&+{32 
u_{r+1} 
\over 3 }
  \sum_{k=0}^{r+1
  }  (5r-5k+11)
  u_k\partial_{u_{k+1}}\partial_{u_{r+2}}^3
\Bigr]\Pi_{\gamma}^{(0)}
\nonumber\\
&+{1\over 768}
\Bigl[
{8\over3}
\sum_{i=0}^{r-3
} [r+1-i]_4 
u_i \partial_{u_{i+2}} \partial_{u_{r+2}}^2 
-64 \sum_{k=0}^{r
} (r-k+3)
u_k  \partial_{u_{k+2}} \partial_{u_{r+2}}^2 
\nonumber\\
&
+{128 u_{r+1}
\over u_{r+2}
}  \sum_{k=0}^{r+1
}(3r-3k+7)
u_k \partial_{u_{k+1}}\partial_{u_{r+2}}^2
\Bigr]\Pi_{\gamma}^{(0)}
\nonumber\\
&-{l(l+1) 
\over12 
}
\Bigl\{
\sum_{k=0}^{r
} u_k (r+2-k) (r-k-3)
\partial_{u_{k+2}}
+{4
u_{r+1}
\over 
u_{r+2} 
}  \sum_{k=0}^{r+1
} (3r-3k+7)
u_k \partial_{u_{k+1}}
\Bigr\}\partial_{u_{r+2}}^2 \Pi_{\gamma}^{(0)}
\nonumber\\
&-{l(l+1)\{1-l(l+1)\} \over 6u_{r+2}}
 \Bigl\{
 {4 u_{r+1} 
 \over u_{r+2}}
 \sum_{k=0}^{r+1
 } u_k (r-k+3)
 \partial_{u_{k+1}}
-\sum_{k=0}^{r
} u_k (r-k+5)
\partial_{u_{k+2}} 
 \Bigr\}
\partial_{u_{r+2}}\Pi_{\gamma}^{(0)}, 
\label{eq:Pi2-exp}
\end{align}
where $[a]_k\equiv a(a-1)\cdots (a-k+1)$ and $u_0=1$.

\subsection{TBA equations and the discontinuity formula}
Since the WKB series (\ref{eq:WKB-series}) is an asymptotic expansion in $\zeta$, we 
define the Borel summation
\be
\hat{\Pi}_{\gamma}(\xi)=\sum_{n\geq0}\frac{1}{(2n)!}\Pi_{\gamma}^{(n)}\xi^{2n}
\ee
and the Borel resummation along the direction $\varphi$
\be
s_{\varphi}
(\Pi_{\gamma})(\hbar)=\frac{1}{\hbar}\int_{0}^{ \infty e^{i\varphi }
}e^{-\xi/\hbar}\hat{\Pi}_{\gamma}(\xi)d\xi.
\ee
In partiular $s(\Pi_{\gamma})(\hbar):=s_{\varphi=0}(\Pi_{\gamma})(\hbar)$ for $\hbar>0$ is the usual Borel resummation.
Here we use the standard notation: $\hbar:=\zeta$. When the  resummed period $s(\Pi_{\gamma})(\hbar)$ converges for small $\hbar$, the WKB period $\Pi_\gamma$ is said to be Borel summable. 
When singularities for  $\hat{\Pi}_{\gamma}(\xi)$ exist  along a direction $\varphi$ in the $\xi$-plane, there arises  a discontinuity for $s_{\varphi}
(\Pi_{\gamma})(\hbar)$, which is defined by
\be
{\rm disc}_\varphi \Pi_\gamma(\hbar)=\lim_{\delta\to 0+}\Big(s(\Pi_\gamma)(e^{i\varphi+i\delta}\hbar)-s(\Pi_\gamma)(e^{i\varphi-i\delta}\hbar)\Big).
\ee
The Delabaere-Pham formula states that the period corresponding to the  classically allowed region  is not Borel summable along the $\varphi=0$ direction, while the  period for the classically forbidden region being Borel summable \cite{dp99}. For the WKB period connecting the simple turning points, the discontinuity formula for classically allowed period is given by \cite{dp99, Iwaki-Nakanishi-2}
\be
\label{eq:DP}
\frac{i}{\hbar}{\rm disc}_{\varphi=0}(\Pi_{\gamma,{\rm allowed}})(\hbar)=\log\prod_{k}\Big(1+\exp\big(-\frac{1}{\hbar}\Pi_{\gamma_{k}}(\hbar)\big)\Big)^{\langle\gamma_{k},\gamma\rangle},
\ee
where $\gamma_k$ is the classically forbidden period, $\langle\gamma_{k},\gamma\rangle$ is the intersection of the cycles. 
When the WKB lines connect a turning point and a simple pole, the discontinuity formula
has been obtained in \cite{Iwaki-Nakanishi-2}.
For example, let us consider the period $\gamma^\prime$ intersecting with $\hat{\gamma}$, such as the $\gamma_1$ in Fig.\ref{fig:r0}.  When this period is classically allowed, the discontinuity formula reads \cite{Iwaki-Nakanishi-2}:
\be
\ba
\frac{i}{\hbar}{\rm disc}_{\varphi=0}(\Pi_{\gamma^\prime,{\rm allowed}})(\hbar)=&\log\prod_{k}\Big(1+\exp\big(-\frac{1}{\hbar}\Pi_{\gamma_{k}}(\hbar)\big)\Big)^{\langle\gamma_{k},\gamma^\prime\rangle}\\
&+\log\Big(1-(e^{2\pi il}+e^{-2\pi il})\exp\big(-\frac{1}{\hbar}\Pi_{\hat{\gamma}}(\hbar)\big)+\exp\big(-2\frac{1}{\hbar}\Pi_{\hat{\gamma}}(\hbar)\big)\Big)^{\langle \hat{\gamma}, \gamma^\prime \rangle }.
\ea
\ee
Moreover, when we rotate $\hbar\to \pm i\hbar$ for real $\hbar$, classically allowed and classically forbidden periods are interchanged, which leads to similar discontinuity formulas for other periods.

Let us go back to our case, where $\Pi_{\gamma_{2i+1}}$ and $\Pi_{\gamma_{2i}}$ are classically allowed and classically forbidden respectively. $\Pi_{\hat{\gamma}}$ is classical allowed for odd $r$, and classically forbidden for even $r$. We then introduce the functions:
\be
\ba
\label{eq:Y-Pi}
&i\log {Y}_{2i+1}^\prime(\theta+\frac{\pi i}{2}\pm i\delta)=\frac{1}{\hbar}s_{\pm\delta}(\Pi_{\gamma{2i+1}})(\hbar),\quad
-\log {Y}_{2i}^\prime(\theta)=\frac{1}{\hbar}\Pi_{2i}(\hbar),\\
&\begin{cases}
i\log\hat{Y}^{\prime}(\theta+\frac{\pi i}{2}\pm i\delta)=\frac{1}{\hbar}s_{\pm\delta}(\Pi_{\hat{\gamma}})(\hbar) & r:{\rm odd}\\
-\log\hat{Y}^{\prime}(\theta)=\frac{1}{\hbar}\Pi_{\hat{\gamma}}(\hbar) & r:{\rm even}
\end{cases}
\ea
\ee
with $\hbar=e^\theta$.  In these functions the discontinuity formula take the unified form:
\be
\ba
-{\rm disc}_{\frac{\pi}{2}}Y_{s}^{\prime}(\theta)&=\log\Big(1+Y_{s-1}\Big)+\log\Big(1+Y_{s-1}\Big),\quad s=1,\cdots,r,\\
-{\rm disc}_{\frac{\pi}{2}}Y_{r+1}^{\prime}(\theta)&=\log\Big(1+Y_{r}(\theta)\Big)+\log\Big(1-(e^{2\pi il}+e^{-2\pi il})Y_{\hat{\gamma}}(\theta)+Y_{\hat{\gamma}}(\theta)^{2}\Big),\\
-{\rm disc}_{\frac{\pi}{2}}\hat{Y}^{\prime}(\theta)&=\log\Big(1+Y_{r+1}(\theta)\Big).
\ea
\ee
These discontinuity formulas together with the large $\theta$ asymptotic behavior of WKB periods, {\it i.e.}
\be
\ba
\label{eq:disc-Y}
-\log Y_s^\prime(\theta)&=m_s e^{\theta}+{\cal O}(e^{-\theta}),\quad s=1,\cdots,r+1,\\
-\log \hat{Y}^\prime(\theta)&=\hat{m} e^{\theta}+{\cal O}(e^{-\theta}),\qquad \theta\to \infty,
\ea
\ee
provides the data for the Riemann-Hilbert problem. The solution to this Riemann-Hilbert problem is given by the TBA equations (\ref{eq:TBA-pole}), from which we could identify $Y^\prime=Y$. To see this in more details, we recall that a pole exists in the kernel of the TBA equations (\ref{eq:TBA-pole}) when ${\rm Im }(\theta'-\theta)=\pm \frac{\pi}{2}$. When the contour of integration crosses this pole, one has to modify the TBA equations to pick up the contribution of the pole. The contribution is easily evaluated by the residue of this pole, which leads the discontinuity formula (\ref{eq:disc-Y}).
From the relation (\ref{eq:Y-Pi}) and comparing the expansion (\ref{eq:Y-exp}) and (\ref{eq:WKB-series}), we find
\be
\ba
\label{eq:m-Pi}
m_{2i+1}^{(n)}&=(-1)^{n}\Pi_{\gamma_{2i+1}}^{(n)},\quad m_{2i}^{(n)}=\Pi_{\gamma_{2i}}^{(n)},\\
\hat{m}^{(n)}&=\begin{cases}
(-1)^{n}\Pi_{\hat{\gamma}}^{(n)} & r:{\rm odd}\\
\Pi_{\hat{\gamma}}^{(n)} & r:{\rm even}
\end{cases}.
\ea
\ee
In the following of this section, we will test our TBA equations by comparing $m_a^{(a)}$ and $\hat{m}^{(n)}$ with the higher-order $\hbar$ correction of WKB periods. More precisely, we will focus on the cases with $r=0$ and $r=1$.

\subsection{$r=0$ case}
Let us consider the case with $r=0$, where the Schr\"odinger equation (\ref{eq:Schrodinger-xu}) 
becomes
\be
\Big(-\hbar^{2}\frac{d^2}{dx^2}+\frac{x^{2}+u_{1}x+u_{2}}{x}+\hbar^2\frac{ l(l+1)}{x^{2}}\Big)\hat{\psi}(x)=0.
\ee
Let $e_i$ ($i=1,2$) be the turning points of the potential $Q_0(x)$, i.e. zeros of $x^2+u_1x+u_2$, which are assumed to be  positive real and ordered as
\be
\label{eq:e1-e2}
0<e_1<e_2,
\ee
$x=0$ is the pole of the potential. In this case, there are two independent WKB cycles: $\gamma_1$ is the cycle connecting $e_1$ and $e_2$, $\hat{\gamma}$ is the cycle connecting $0$ and $e_1$ (Fig.\ref{fig:r0}).

The classical WKB period  $\Pi^{(0)}$ can be evaluated explicitly by using the hypergeometric function
\be
\ba
\Pi^{(0)}_{\gamma_1}=\frac{2}{i}\int_{e_{1}}^{e_{2}}\sqrt{\frac{x^{2}+u_{1}x+u_{2}}{x}}dx=2\frac{(e_{1}-e_{2})^{2}}{\sqrt{e_{1}}}\frac{\Gamma(3)}{\Gamma(\frac{1}{2})\Gamma(\frac{3}{2})}F\big(\frac{1}{2},\frac{3}{2},3,\frac{e_{1}-e_{2}}{e_{1}}\big),
\ea
\ee
\be
\ba
{\Pi}^{(0)}_{\hat{\gamma}}=2\int_{0}^{e_{1}}\sqrt{\frac{x^{2}+u_{1}x+u_{2}}{x}}dx=2e_{1}\sqrt{e_{2}}\frac{\Gamma(2)}{\Gamma(-\frac{1}{2})\Gamma(\frac{1}{2})}F\big(-\frac{1}{2},\frac{1}{2},2,\frac{e_{1}}{e_{2}}\big).
\ea
\ee
Here $F(a,b;c;z)$ is the hypergeometric function whose integral representation is given by\footnote{We followed the notation of \cite{Masuda:1996xj}, which is different with the notation of ${{}_2}F_1$ in Mathematica.}
\be
F(a,b;c;x)=\frac{\Gamma(a)\Gamma(b)}{\Gamma(c)}\int_{0}^{1}t^{b-1}(1-t)^{c-b-1}(1-tx)^{-a} dt.
\ee
The corrections to the classical WKB periods $\Pi_\gamma^{(1)}$ and $\Pi_\gamma^{(2)}$  can be evaluated by using the formulas (\ref{eq:Pi1-exp}) and (\ref{eq:Pi2-exp}) respectively.

On the other hand, we substitute  the masses $m_1$ and $\hat{m}$ given by (\ref{eq:logyasymp}) into the TBA equations, which are given by
\be
\ba
\label{eq:TBA-r0}
\log Y_{1}(\theta)&=-m_{1}e^{\theta}+\int_{\mathbb{R}}\frac{\log\Big(1+\omega^{\frac{3}{2}} e^{2\pi il}\hat{Y}(\theta^{\prime})\Big)}{\cosh(\theta-\theta^{\prime})}\frac{{d}\theta^{\prime}}{2\pi}+\int_{\mathbb{R}}\frac{\log\Big(1+\omega^{\frac{3}{2}}e^{-2\pi il}\hat{Y}(\theta^{\prime})\Big)}{\cosh(\theta-\theta^{\prime})}\frac{{d}\theta^{\prime}}{2\pi},\\
\log\hat{Y}(\theta)&=-\hat{m}e^{\theta}+\int_{\mathbb{R}}\frac{\log\Big(1+Y_{1}(\theta^{\prime})\Big)}{\cosh(\theta-\theta^{\prime})}\frac{{d}\theta^{\prime}}{2\pi}.
\ea
\ee
Solving the TBA equations numerically, we evaluate the coefficients (\ref{eq:mn}) of the large $\theta$ expansions of the Y-functions.

In Table \ref{tab:r0-TBAvsWKB}, we compare $m_{1}^{(n)}$ and $\hat{m}^{(n)}$ ($n=1,2$) to 
the quantum periods $\Pi_{\gamma_{1}}^{(n)}$ and $\hat{\Pi}^{(n)}$ with fixed $u_1$, $u_2$ and $l$. As we see in this Table, the two calculations agree to the relation (\ref{eq:m-Pi}) in high numeric precision. In the case where $-1\leq l\leq 0$, i.e. the basis of the solutions $x^{-l}+\cdots$ and $x^{-l-1}+\cdots$ are regular at $x=0$, we have tested our TBA equations with high numerical precision.

\begin{table}[tb]
\begin{center}
  \begin{tabular}{ccccc}\hline
	$n$ &  $\Pi_{\gamma_1}^{(n)} $                 &  $m_1^{(n)} $             &  $\Pi_{\hat{\gamma}}^{(n)}$     &  $\hat{m}^{(n)} $         \\ \hline
    $0$  & $3.4222512870$ & $3.4222512870$ & $1.9055241299$ & $1.9055241299$\\
	$1$  & $0.2960549383$  & $-0.2960549173$  & $-0.1166722969$  & $-0.1166722826$        \\
	$2$ &$0.1580844621$   & $0.1580844602$  & $0.01972364328$   & $0.01972364274$    \\ \hline
 \end{tabular}
  \caption{The higher-order corrections of the WKB periods for $r=0$ case with $u_1=-3$, $u_2=1$ and $l=-2/5$. The numerical calculation of the TBA equations is performed by the Fourier discretization with $2^{12}$ points in the region $(-L,L)$ with the cutoff $L=25$.}
  \vspace{3mm}\label{tab:r0-TBAvsWKB}
\end{center}
\end{table}

\subsubsection{Effective central charge and PNP relation}
The formula for the effective central charge (\ref{eq:ceff-diag}) can be rewritten in terms of the masses and their next order correction (\ref{eq:mn}):
\be
c_{\rm eff}=-\frac{6}{\pi}(m_1\hat{m}^{(1)}+\hat{m}m_1^{(1)})=2\big(1-8(l+\frac{1}{2})^2\big).
\ee
This leads to the relations for the quantum periods
\be
\Pi_{\gamma_1}^{(0)}{\Pi}^{(1)}_{\hat{\gamma}}-{\Pi}_{\hat{\gamma}}^{(0)}\Pi_{\gamma_1}^{(1)}=-\frac{\pi}{3}\big(1-8(l+\frac{1}{2})^2\big).
\label{eq:pnpr0}
\ee
which is known as the PNP relations or a quantum version of the Matone relations \cite{Matone:1995rx, Sonnenschein:1995hv,Eguchi:1995jh, Dunne:2014bca,Gorsky:2014lia,Basar:2015xna,Codesido:2016dld,Basar:2017hpr}.

We consider (\ref{eq:pnpr0}) in  the limit $u_1\rightarrow 0$, where the inverse potential term vanishes. 
In this case, one of the turning points goes to zero.  Then the classical period $\Pi_{\hat{\gamma}}^{(0)}$ will vanish. However, the $\hbar^2$-order correction $\Pi_{{\gamma}_1}^{(1)}$ becomes diverge in this limit due to the second term in (\ref{eq:Pi1-exp}).  The l.h.s of (\ref{eq:pnpr0}) remains finite in this limit.
We have numerically tested our TBA equations  with vanishing $\hat{m}$ against the next order correction of $\hat{Y}$. 
We thus found a precise match numerically. Moreover, from the numerical solution, the  effective central charge (\ref{eq:ceff-massless}) is evaluate as
\be
c_{{\rm eff}}|_{u_2=0}=1-\frac{12(l+\frac{1}{2})^{2}}{3},
\ee
which reduces to the one in Appendix D.2 of \cite{Dorey:2007zx}.

\subsubsection{Voros spectrum in Bohr-Sommerfeld approximation}
Let us now compute the spectrum of Schr\"odinger equation by using the TBA equations. The discontinuity formula has shown that the classically allowed period $\Pi_{\gamma_1}$ is not Borel summable and need to be resummed for positive and real $\hbar$. In the context of TBA, the Borel non-summability of $\Pi_{\gamma_1}$ means that we will hit a pole after shifting $\theta\to \theta+\frac{\pi i}{2}$ in $\log Y_1(\theta)$. In this paper, we resumme the period by taking the average of two lateral resummations above and below the singular point, namely ``median'' resummation denoted by $s_{\rm med}(\Pi_{\gamma_1})$.
 In the TBA equations, this resummation can be written as
\be
\label{eq:smedPi1}
\frac{1}{\hbar}s_{{\rm med}}(\Pi_{\gamma_{1}})(\hbar)=m_{1}e^{\theta}+{\rm P}\int_{\mathbb{R}}\frac{\log\Big(1+\omega^{\frac{3}{2}}e^{2\pi il}\hat{Y}(\theta^{\prime})\Big)\Big(1+\omega^{\frac{3}{2}}e^{-2\pi il}\hat{Y}(\theta^{\prime})\Big)}{\sinh(\theta-\theta^{\prime})}\frac{d\theta^{\prime}}{2\pi},
\ee
where $\hbar=e^{-\theta}$ and ${\rm P}$ is the principal value of the singular integral computed by
\be
{\rm P}\int_{\mathbb{R}}\frac{f(\theta^{\prime})}{\sinh(\theta-\theta^{\prime})}d\theta^{\prime}=\lim_{\delta\to0}\int_{\mathbb{R}}\frac{\sinh(\theta-\theta^{\prime})\cos(\delta)}{\sinh^{2}(\theta-\theta^{\prime})\cos^{2}(\delta)+\cosh^{2}(\theta-\theta^{\prime})\sin^{2}(\delta)}f(\theta^{\prime})d\theta^{\prime}.
\ee
The exact quantization condition for this problem is not known so far. Here, we consider the Bohr-Sommerfeld (BS) quantization condition:
\be
\label{eq:BSQC-r=0}
\frac{1}{\hbar}s_{{\rm med}}(\Pi_{\gamma_{1}})(\hbar)\sim 2\pi (k+\frac{1}{2}),\quad k\in \mathbb{Z}_{\geq 0}. 
\ee
\begin{figure}[tbh]
\begin{center}
\resizebox{75mm}{!}{\includegraphics{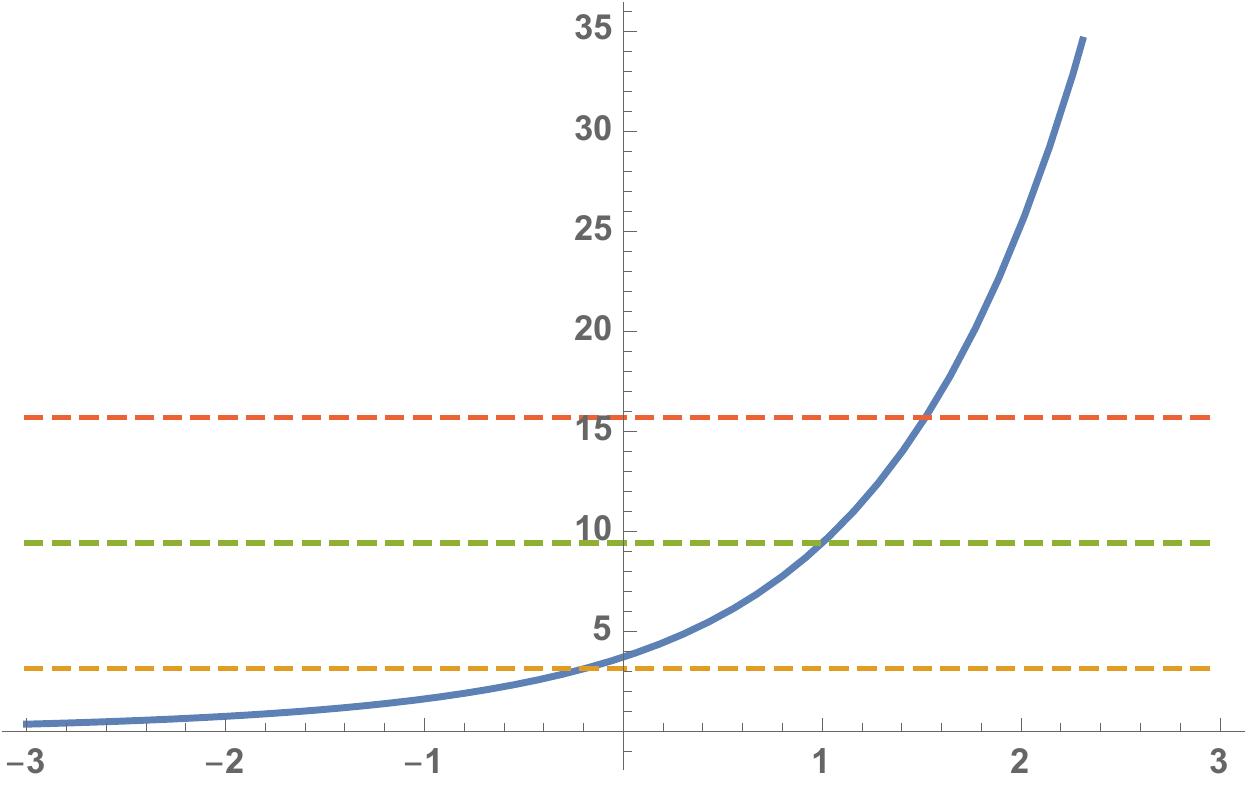}}
\end{center}
  \caption{The profile of the median resummation of the WKB period $s_{\rm med}(\Pi_{\gamma_1})$ as a function of $\theta$ for $u_1=-3$, $u_2=1$ and $l=-2/5$. The horizontal dash lines are $\pi$, $3\pi$ and $5\pi$. We used 
  a discretization with $2^{12}$ points and the cutoff $L=50$.}
\label{fig:VS0}
\end{figure}
In Fig.\ref{fig:VS0}, we plot the median resummation of WKB period $\Pi_{\gamma_1}$ for $u_1=-3$, $u_2=1$ and $l=-2/5$, obtaining from the TBA equations. The horizontal dash lines are $\pi$, $3\pi$ and $5\pi$. The intersect points provide the first three values of Voros spectrum, i.e. the spectrum of $e^{-\theta}=\hbar$ with fixed moduli. In Table \ref{tab:VSr0}, we present the Voros spectrum computed from the BS quantization condition (\ref{eq:BSQC-r=0}) combined with the TBA equations and the one obtained from the WKB approximation (the BS quantization condition combined with the classical periods). Our numerical result show that the Voros spectrum becomes smaller when level $k$ increases. This means that for  large  $k$, where $\hbar$ will be small enough, the driving term  $m_1e^\theta$ in (\ref{eq:smedPi1})  mainly contributes to the periods.
Then the WKB approximation becomes valid for higher-level $k$, which is confirmed by numerical 
calculations.

\begin{table}[tb]
\begin{center}
  \begin{tabular}{cccc}\hline
	Level & $e^{-\theta_{\rm TBA}}$ & $e^{-\theta_{\rm WKB}}$  \\ \hline
	 $0$   & $1.216954845$  & $1.089336418$                \\
	 $1$   & $0.367759415$  &$0.363112139$     \\
	 $2$   & $0.218798031$ & $0.217867284$ \\  
	 $3$   & $0.155951604$ & $0.155619488 $ \\ \hline
  \end{tabular}
  \caption{The first column shows the spectrum of values of $e^{-\theta}$ for $u_1=-3$, $u_2=1$ and $l=-2/5$, as computed from the BS quantization condition (\ref{eq:BSQC-r=0}) and the TBA equations (\ref{eq:TBA-r0}). The second column shows the spectrum of values of $e^{-\theta}$, as computed from the BS quantization condition (\ref{eq:BSQC-r=0}) and the WKB approximation.}
  \vspace{3mm}
  \label{tab:VSr0}
\end{center}
\end{table}

\subsection{$r=1$ case}
We next consider $r=1$ case, where the ODE (\ref{eq:Schrodinger-xu}) becomes
\be
\Big(-\hbar^{2}\partial_{x}^{2}+\frac{x^{3}+u_{1}x^2+u_{2}x+u_3}{x}+\hbar^2 \frac{l(l+1)}{x^{2}}\Big)\hat{\psi}(x)=0.
\ee
Let us choose the parameters $u_1$, $u_2$ and $u_3$ such that the zeros $e_i$($i=1,2,3$) of $x^3+u_1 x^2+u_2 x+u_3$ are positive real and ordered as
\be
0<e_1<e_2<e_3.
\ee
In this case, there are three independent WKB cycles. $\gamma_1$ is the cycle connecting $e_2$ and $e_3$. $\gamma_2$  is the cycle connecting $e_1$ and $e_2$. $\hat{\gamma}$ is the cycle connecting the pole $0$ and $e_1$(Fig.\ref{fig:r1}).
In this case, all the associated classical WKB periods are positive and real.

Let us now compare the higher-order correction of WKB periods. The $u_3$-derivative of the classical period $\Pi^{(0)}_\gamma$ can be expressed in terms of the hypergeometric function:
 \be
  \ba
  \partial_{u_{3}}\Pi_{\gamma_1}^{(0)}&=\frac{1}{i}\int_{e_{2}}^{e_{3}}dx\frac{1}{\sqrt{x^{4}+u_{1}x^{3}+u_{2}x^{2}+u_{3}x}}\\
  &=\frac{1}{i}\frac{1}{\sqrt{e_{1}e_{3}-e_{2}e_{3}}}\frac{\Gamma(1)}{\Gamma(\frac{1}{2})\Gamma(\frac{1}{2})}F(\frac{1}{2},\frac{1}{2};1;\frac{e_{1}e_{3}-e_{1}e_{2}}{e_{1}e_{3}-e_{2}e_{3}}),
  \ea
  \ee

\be
\ba
\partial_{u_{3}}\Pi_{\gamma_2}^{(0)}
&=\int_{e_{1}}^{e_{2}}dx\frac{1}{\sqrt{x(x-e_{1})(x-e_{2})(x-e_{3})}}\\
&=\frac{\Gamma(1)}{\Gamma(\frac{1}{2})\Gamma(\frac{1}{2})}\frac{1}{\sqrt{-e_{1}e_{2}+e_{2}e_{3}}}F(\frac{1}{2},\frac{1}{2};1;\frac{e_{1}e_{3}-e_{2}e_{3}}{e_{1}e_{2}-e_{2}e_{3}}),
\ea
\ee

\be
\ba
\partial_{u_{3}}{\Pi}_{\hat{\gamma}}^{(0)}&=\frac{1}{i}\int_{0}^{e_{1}}\frac{1}{\sqrt{x^{4}+u_{1}x^{3}+u_{2}x^{2}+u_{3}x}}dx\\
&=\frac{1}{i}\frac{1}{\sqrt{e_{1}e_{3}-e_{2}e_{3}}}\frac{\Gamma(1)}{\Gamma(\frac{1}{2})\Gamma(\frac{1}{2})}F(\frac{1}{2},\frac{1}{2};1;\frac{e_{1}e_{3}-e_{1}e_{2}}{e_{1}e_{3}-e_{2}e_{3}}).
\ea
\ee
From these equations and (\ref{eq:Pi1-exp}),  (\ref{eq:Pi2-exp}), we can evaluate explicitly 
 $\Pi^{(1)}_\gamma$ and  $\Pi_{\gamma}^{(2)}$.

On the other hand, the TBA system is given by
\be
\ba
\label{eq:TBA-r1}
\log Y_{1}(\theta)&=-m_{1}e^{\theta}+\int_{\mathbb{R}}\frac{\log\Big(1+Y_{2}(\theta^{\prime})\Big)}{\cosh(\theta-\theta^{\prime})}\frac{{d}\theta^{\prime}}{2\pi},\\
\log Y_{2}(\theta)&=-m_{1}e^{\theta}+\int_{\mathbb{R}}\frac{\log\Big(1+Y_{1}(\theta^{\prime})\Big)}{\cosh(\theta-\theta^{\prime})}\frac{{d}\theta^{\prime}}{2\pi}+\int_{\mathbb{R}}\frac{\log\Big(1+\omega^{2}e^{2\pi il}\hat{Y}(\theta^{\prime})\Big)}{\cosh(\theta-\theta^{\prime})}\frac{{d}\theta^{\prime}}{2\pi}\\
&\qquad\ \qquad+\int_{\mathbb{R}}\frac{\log\Big(1+\omega^{2}e^{-2\pi il}\hat{Y}(\theta^{\prime})\Big)}{\cosh(\theta-\theta^{\prime})}\frac{{d}\theta^{\prime}}{2\pi},\\
\log\hat{Y}(\theta)&=-\hat{m}e^{\theta}+\int_{\mathbb{R}}\frac{\log\Big(1+Y_{2}(\theta^{\prime})\Big)}{\cosh(\theta-\theta^{\prime})}\frac{{d}\theta^{\prime}}{2\pi}.
\ea
\ee
where $m_1$, $m_2$ and $\hat{m}$ are  real and positive.
In Table \ref{tab:r1-TBAvsWKB}, we compare $m_{1}^{(n)}$ and $ m_{1}^{(n)}$, $n=1,2$, calculated numerically from (\ref{eq:mn}) by using the TBA equations, to 
the quantum periods $\Pi_{\gamma_{1}}^{(n)}$ and $ \Pi_{\gamma_{2}}^{(n)}$ with some fixed values of $u_1$, $u_2$, $u_3$ and $l$. 
We find that the two calculations 
numerically agree with each other and the relation (\ref{eq:m-Pi}) is confirmed. 
Moreover, eq. (\ref{eq:mn}) predicts that $m_1^{(n)}=\hat{m}^{(n)}$ for all $n>0$, which is also 
confirmed for $n=1,2$ by numerical calculations.
This implies the quantum periods must satisfy   $\Pi_{\gamma_1}^{(n)}=\Pi_{\hat{\gamma}}^{(n)}$ for $n>0$, which is obvious for $n=1,2$ because of $\partial_{u_3}\Pi^{(0)}_{\gamma_1}=\partial_{u_3}\Pi^{(0)}_{\hat{\gamma}}$.
\begin{table}[tb]
\begin{center}
  \begin{tabular}{ccccccc}\hline
	$n$ &  $\Pi_{\gamma_1}^{(n)} $                 &  $m_1^{(n)} $             &  $\Pi_{{\gamma}_2}^{(n)}$  &  ${m}_2^{(n)} $ &$\Pi_{\hat{\gamma}}^{(n)} $                 &  $\hat{m}^{(n)} $        \\ \hline
	$0$  &$6.40730315$ &$6.40730315$& $0.849415273$&$0.849415273$&$4.64015728$&$4.64015728$\\
	$1$  & $0.2954268$  & $-0.29542654$  & $0.0602544$  & $0.0602547$ & $0.2954268$  & $-0.29542654$       \\
	$2$ &$0.9636413$   & $0.9636412$  & $-0.009866916$   & $-0.009866912$ &$0.9636413$   & $0.9636412$   \\ \hline
  \end{tabular}
  \caption{The higher-order corrections of the WKB periods for $r=1$ case with $u_1=-13/2$, $u_2=10,u_3=-4$ and $l=-1/10$. The numerical calculation of the TBA equations is performed by Fourier discretization with $2^{12}$ points of the region $(-L,L)$ with the cutoff $L=25$.}
  \vspace{3mm}\label{tab:r1-TBAvsWKB}
\end{center}
\end{table}
We have tested our TBA equations with higher numeric precision, when $-1\leq l\leq 0$, i.e. the basis $x^{-l}$ and $x^{-l-1}$ at $x\to 0$ are regular.

\subsubsection{Effective central charge and PNP relation}
For $r=1$, the effective central charge (\ref{eq:ceff-diag}) can be rewritten as 
\be
c_{\rm eff}=-\frac{6}{\pi}(m_2\hat{m}^{(1)}+\hat{m}m_2^{(1)})+\frac{6}{\pi^2}(m_1-\hat{m})\int_{\mathbb{R}} e^\theta
\log\Big(1+Y_1(\theta)\Big)d\theta.
\ee
In particular, when $m_1=\hat{m}$ ($\Pi_{\gamma_1}^{(0)}=\Pi_{\hat{\gamma}}^{(0)}$), this leads to the PNP relation: 
\be
{\Pi}_{\hat{\gamma}}^{(0)}\Pi_{\gamma_2}^{(1)}-\Pi_{\gamma_2}^{(0)}{\Pi}^{(1)}_{\hat{\gamma}}=-\frac{\pi}{2}\big(1-6(l+\frac{1}{2})^2\big).
\ee

We then consider the case in which $u_3=0$,
where 
one of the turning points approaches to zero. The classical period of $\hat{Y}$ will vanish. The numerical solution of effective central charge (\ref{eq:ceff-massless}) is evaluate as
\be
c_{{\rm eff}}\big|_{u_3= 0}=2-6(l+\frac{1}{2})^{2}.
\ee
Note that taking the limit $l\to 0$, we obtain the effective central charge $c_{\rm eff}\to \frac{1}{2}$, which coincides with the $r=1$ limit of the effective central charge obtained in \cite{Ito:2018eon}.

\subsubsection{Voros spectrum in Bohr-Sommerfeld approximation}
For $r=1$,  the period $\Pi_{\gamma_1}$ is non-Borel summable.
In the TBA equations, the median resummation of the period is done by
\be
\frac{1}{\hbar}s_{{\rm med}}(\Pi_{\gamma_{1}})(\hbar)=m_{1}e^{\theta}+{\rm P}\int_{\mathbb{R}}\frac{\log\Big(1+Y_2(\theta^{\prime})\Big)}{\sinh(\theta-\theta^{\prime})}\frac{d\theta^{\prime}}{2\pi}.
\ee

In Table \ref{tab:VSr1}, we calculate the Voros spectrum $e^{-\theta}$ from the TBA equations by using the BS quantization condition and compared them with those of the WKB approximation.
The numerical results are similar to the $r=0$ results.
\begin{table}[tb]
\begin{center}
  \begin{tabular}{cccc}\hline
	Level & $e^{-\theta_{\rm TBA}}$ & $e^{-\theta_{\rm WKB}}$  \\ \hline
	 $0$   & $2.110949740$  & $2.039507935$                \\
	 $1$   & $0.690146946$  &$0.679835978$     \\
	 $2$   & $0.411127634$ & $0.407901587$ \\  
	 $3$   & $0.292665832$ & $0.291358276$ \\ \hline
  \end{tabular}
  \caption{The first column shows the spectrum of values of $e^{-\theta}$ for $r=1$ case with $u_1=-13/2$, $u_2=10,u_3=-4$ and $l=-1/10$, as computed from the BS quantization condition and the TBA equations. The second column shows the spectrum of values of $e^{-\theta}$, as computed from the BS quantization and the WKB approximation.}
  \vspace{3mm}
  \label{tab:VSr1}
\end{center}
\end{table}


\section{Conclusions and discussions}
In this paper, we have derived a set of TBA equations from the Schr\"odinger equation with polynomial potential and a term with a regular singularity.
This provides further non-trivial examples for the ODE/IM correspondence. 
The TBA equations provide solutions for the Riemann-Hilbert problem of the WKB periods, and govern the exact $\hbar$-dependence of the WKB periods.
From the TBA equations, We have introduced the effective central charge, which coincides the one defined in \cite{Ito:2018eon} by taking the limit $l\to 0$ and $u_{r+2}\to 0$. The effective central charge also leads to non-trivial constraints on the classical WKB periods and the first-order correction of the WKB periods, which constrains are related to the PNP relations. As an application, we compute numerically the Voros spectrum using  the Bohr-Sommerfeld quantization condition.

There are many open questions. In the present paper, we have assumed all the turning points take different values. It is interesting to consider the Schr\"odinger equations with degenerate potential such as $(x^{m}-u)^{K}$, from which one may derive the Hybrid NLIEs in \cite{Suzuki:1998ve,Dunning:2002tt,Suzuki:2004zw,Babenko:2017fmu}. It is also interesting to consider the Schr\"odinger equation with so-called the monster potential, which corresponds to the excited states of the quantum integral model 
\cite{Bazhanov:2003ni,Fioravanti:2004cz,Masoero:2018rel,Bazhanov:2019xvy,Kotousov:2019ygw}.
So far, we have worked out the Schr\"odinger equation with only one irregular singular point. It is very interesting and important to consider two irregular singular points in the potential, which helps us to study the ${\cal N}=2$ four-dimensional $SU(N)$ super Yang-Mills theory \cite{Zam-GME, Grassi:2018bci,poster,Hollands:2019wbr,Grassi:2019coc,Fioravanti:2019vxi, Fioravanti:2019awr} and the non-planar scattering amplitude in ${\cal N}=4$ super Yang-Mills theory from the AdS side \cite{Ben-Israel:2018ckc}. The other interesting direction is to study the higher-order ODEs with polynomial potentials and poles, which are related to the quantum Seiberg-Witten curve of Argyres-Douglas theories in \cite{Ito:2017ypt,Ito:2019twh}.

There are many ODEs are essentially equivalent after the coordinate transformation or parameter redefinitions. One may derive different type TBA equations from these equivalent ODEs. In the appendix C, we have seen these different type TBA equations are  essentially equivalent, through a simple but nontrivial example. It is very important to find the general transform rule of the TBA equations under the coordinate transform, which helps us to derive the TBA equations for the ODE with more general potential.

\section*{Acknowledgements}
We would like to thank Takayasu Kondo, Kohei Kuroda, Marcos Marino, Ryo Suzuki, Roberto Tateo and Dmytro Volin for useful discussions.
H.S. would like to thank Jilin University, South China Normal University, Sun Yat-Sen University and Korea Institute for Advanced Study for the warm hospitality.
The work of K.I. is supported in part by Grant-in-Aid for Scientific Research 18K03643 and  17H06463 from Japan Society for the Promotion of Science (JSPS). The work of H.S. is supported by the grant ``Exact Results in Gauge and String Theories'' from the Knut and Alice Wallenberg foundation.


\appendix

\section{T-Q relation and Bethe ansatz equation}
\label{sec:TQ-BAE}
In this appendix, we derive the T-Q relation from the second order differential equation (\ref{eq:Schrodinger-eq-zb}). Let $y_k$ ($k\in {\mathbb Z}$) be a set of solutions obtained by Symanzik rotation.
Since $(y_0,y_1)$ form a basis of the solutions,  $y_k$ can be expanded as
\be
\label{eq:yk-y0y1expansion}
y_{k}=\frac{W_{k,1}}{W_{0,1}}y_{0}+\frac{W_{0,k}}{W_{0,1}}y_{1}.
\ee
In particular, $y_{-1}$ is expanded as
\be
\label{eq:y-1-expand}
y_{-1}=\frac{W_{-1,1}}{W_{0,1}}y_{0}-\frac{W_{-1,0}}{W_{0,1}}y_{1}.
\ee
By using the basis $(\psi_+,\psi_-)$ around $z=0$ defined in (\ref{eq:basis-z=0}), one can expand $y_0$ in this basis as
\be
(2l+1)y_{0}=Q_{+}\psi_{-}-Q_{-}\psi_{+},
\label{eq:y0}
\ee
where $Q_+$ and $Q_-$ are the Q-functions. Taking the Wronskian with $y_0$ and $\psi_{\pm}$, we find
\be
\label{eq:Q-def}
Q_{\pm}(b_{a})=W[y_{0},\psi_{\pm}](b_{a}).
\ee
By the Symanzik rotation,  $y_1$ is expressed as
\be
y_{1}=-\omega^{-(l+\frac{1}{2})}\frac{Q_{-}(\omega^{-a}b_{a})}{2l+1}\psi_{+}+\omega^{(l+\frac{1}{2})}\frac{Q_{+}(\omega^{-a}b_{a})}{2l+1}\psi_{-}.
\label{eq:y1}
\ee
From (\ref{eq:y0}) and (\ref{eq:y1}), we obtain
\be
\left(\begin{array}{c}
y_{1}\\
y_{0}
\end{array}\right)(z)=
{\cal Q}(b_{a})\left(\begin{array}{c}
\psi_{+}\\
\psi_{-}
\end{array}\right)(z), 
\qquad 
{\cal Q}(b_{a}):=\left(\begin{array}{cc}
-\omega^{-(l+\frac{1}{2})}\frac{Q_{-}(\omega^{-a}b_{a})}{2l+1} & \omega^{(l+\frac{1}{2})}\frac{Q_{+}(\omega^{-a}b_{a})}{2l+1}\\
-\frac{Q_{-}(b_{a})}{2l+1} & \frac{Q_{+}(b_{a})}{2l+1}
\end{array}\right).
\ee
From the Symanizk rotation, we also find
\be
\ba
\label{eq:yk-psi-pm}
W[y_{k},\psi_{+}](b_{a})&=\omega^{k(l+\frac{1}{2})}W[y,\psi_{+}](\omega^{-ak}b_{a}), \\
W[y_{k},\psi_{-}](b_{a})&
=\omega^{-k(l+\frac{1}{2})}W[y,\psi_{-}](\omega^{-ak}b_{a}).
\ea
\ee
Let us take the Wronskian of $y_{-1}$ in (\ref{eq:y-1-expand}) with $\psi_\pm(b_a)$.
The result is
\be
W[y_{-1},\psi_{\pm}]=\frac{W_{-1,1}}{W_{0,1}}W[y_{0},\psi_{\pm}]-\frac{W_{-1,0}}{W_{0,1}}W[y_{1},\psi_{\pm}].
\ee
Using (\ref{eq:Q-def}) and (\ref{eq:yk-psi-pm}), we obtain the T-Q relation
\be
\label{eq:T-Q-relation}
\frac{W_{-1,1}}{W_{0,1}}Q_{\pm}(b_{a})=\omega^{\mp(l+\frac{1}{2})}Q_{\pm}(\omega^{a}b_{a})+\frac{W_{-1,0}}{W_{0,1}}\omega^{\pm(l+\frac{1}{2})}Q_{\pm}(\omega^{-a}b_{a}),
\ee
where $\frac{W_{-1,1}}{W_{0,1}}$ is regarded as the T-function. 
If we set the zeros of $Q_a$ as $b_{a,k}$ ($k\in \mathbb{Z}$):
\be
Q_+(b_{a,k})=0 ,
\ee
then (\ref{eq:T-Q-relation}) becomes
\be
\frac{Q_{+}(\omega^{a}b_{a,k})}{Q_{+}(\omega^{-a}b_{a,k})}=-\frac{W_{-1,0}}{W_{0,1}}\omega^{2(l+\frac{1}{2})}.
\ee
In our normalization (\ref{eq:normalization}), we obtain the Bethe ansatz equation of the form:
\be
\frac{Q_{+}(\omega^{a}b_{a,k})}{Q_{+}(\omega^{-a}b_{a,k})}=\begin{cases}
-\omega^{2(l+\frac{1}{2})} & r+1:{\rm odd}\\
-\omega^{-2B_{\frac{r+3}{2}}}\omega^{2(l+\frac{1}{2})} & r+1:{\rm even}
\end{cases}.
\ee


\section{Wall crossing of TBA equations}\label{app:WC}
In section 4, we note that the TBA equations (\ref{eq:TBA-com-mass}) are valid in the minimal chamber (\ref{eq:minimal-chamber}) in the moduli space. In this appendix, we explain how to analytically continue the TBA equations (\ref{eq:TBA-com-mass}) to outside of the minimal chamber \cite{Alday:2010vh, Toledo10, Toledo16}. Here we consider two particular cases: 
1) $\frac{\pi}{2}<\phi_2 -\phi_1<\pi$ while all other difference of phases $\phi$ are in between $-\pi/2$ and $\pi/2$, 
2) $\frac{\pi}{2}<\phi_{r+1}-\hat{\phi}<\pi$ while all other difference of phases $\phi$ are in between $-\pi/2$ and $\pi/2$. 
Any other examples of the wall crossing can be treated similarly.

\subsection{Case 1: $\phi_2 -\phi_1$ crosses $\pi/2$}
When the differences of the phases $\phi_2 -\phi_1$ crosses the value of $\frac{\pi}{2}$ while  all other differences of the phases $\phi_s$'s and $\hat{\phi}$ are in between $-\pi/2$ and $\pi/2$, we need to modify the integration contour in the TBA equations which results in  including the contribution of the pole in the kernel.
The TBA system is then modified as
\be
\ba
\label{eq:Y1Y2-WC}
\log Y_{1}(\theta-i\phi_{1})&=-|m_{1}|e^{\theta}+\log\Big(1+Y_{2}(\theta-i\phi_{1}-\frac{i\pi}{2}+i\delta)\Big)\\&\quad+\int_{\mathbb{R}}\frac{\log\Big(1+Y_{2}(\theta^{\prime}-i\phi_{2})\Big)}{\cosh(\theta-\theta^{\prime}-i\phi_{1}+i\phi_{2})}\frac{d\theta^{\prime}}{2\pi}.\\
\log Y_{2}(\theta-i\phi_{2})&=-|m_{2}|e^{\theta}+\log\Big(1+Y_{1}(\theta-i\phi_{2}+\frac{i\pi}{2}-i\delta)\Big)\\
&\quad+\int_{\mathbb{R}}\frac{\log\Big(1+Y_{1}(\theta^{\prime}-i\phi_{1})\Big)}{\cosh(\theta-\theta^{\prime}-i\phi_{2}+i\phi_{1})}\frac{d\theta^{\prime}}{2\pi}+\int_{\mathbb{R}}\frac{\log\Big(1+Y_{3}(\theta^{\prime}-i\phi_{3})\Big)}{\cosh(\theta-\theta^{\prime}-i\phi_{2}+i\phi_{3})}\frac{d\theta^{\prime}}{2\pi}.
\ea
\ee
while other TBA equations in (\ref{eq:TBA-com-mass}) are unchanged.  However, the arguments of the Y-functions in the second terms of right hand side are different the ones on the left hand side. 
To obtain a closed TBA system, we need to add the Y-functions with the argument being shifted. The additional TBA equations are
\be
\ba
\label{eq:Y1Y2-shifted}
\log Y_{1}(\theta-i\phi_{2}+\frac{i\pi}{2}-i\delta)&=-|m_{1}|e^{\theta+i\phi_{1}-i\phi_{2}+\frac{i\pi}{2}-i\delta}+\log\Big(1+Y_{2}(\theta-i\phi_{2})\Big)\\
&\quad+\int_{\mathbb{R}}\frac{\log\Big(1+Y_{2}(\theta^{\prime}-i\phi_{2})\Big)}{\cosh(\theta-\theta^{\prime}+\frac{i\pi}{2}-i\delta)}\frac{d\theta^{\prime}}{2\pi},\\
\log Y_{2}(\theta-i\phi_{1}-\frac{i\pi}{2}+i\delta)&=-|m_{2}|e^{\theta+i\phi_{2}-i\phi_{1}-\frac{i\pi}{2}+i\delta}+\log\Big(1+Y_{1}(\theta-i\phi_{1}+i\delta)\Big)\\
&\quad+\int_{\mathbb{R}}\frac{\log\Big(1+Y_{1}(\theta^{\prime}-i\phi_{1})\Big)}{\cosh(\theta-\theta^{\prime}-\frac{i\pi}{2}+i\delta)}\frac{d\theta^{\prime}}{2\pi}+\int_{\mathbb{R}}\frac{\log\Big(1+Y_3(\theta^\prime -i\phi_3)}{\cosh(\theta-\theta^{\prime}-i\phi_{1}+i\phi_3-\frac{i\pi}{2}+i\delta)}\frac{d\theta^{\prime}}{2\pi}.
\ea
\ee
The TBA equations (\ref{eq:Y1Y2-WC}) and (\ref{eq:Y1Y2-shifted}) together with the TBA equations with $a=3,\cdots,r+1$ in (\ref{eq:TBA-com-mass}) provide a closed TBA system.

One may absorb the extra terms $\log\Big(1+Y_{1,2}(\theta-\phi_{2,1}\pm \frac{\pi}{2})\Big)$ on the right hand side of (\ref{eq:Y1Y2-WC}) to left hand side, and denote them as new Y-functions. As a result, we will obtain a closed TBA system with $r+3$ Y-functions.
This type of wall-crossing of the TBA equations is similar to that of \cite{Alday:2010vh,Ito:2018eon} .

\subsection{Case 2: $\phi_{r+1}-\hat{\phi}$ crosses $\pi/2$}

When the phase $\phi_1-\hat{\phi}$ crosses the value of $\pi/2$, while all other differences of phases are in $(-{\pi\over2},{\pi\over2})$, we need to modify the TBA equations to 
\be
\ba
\label{eq:TBA-WC-Yr+1-Yhat}
\log Y_{r+1}(\theta-i\phi_{1})&=-|m_{r+1}|e^{\theta}+\int_{\mathbb{R}}\frac{\log\Big(1-2\cos(2\pi l)\hat{Y}(\theta^{\prime}-\hat{\phi})+\hat{Y}(\theta^{\prime}-\hat{\phi})^{2}\Big)}{\cosh(\theta-\theta^{\prime}-i\phi_{r+1}+i\hat{\phi})}\frac{d\theta^{\prime}}{2\pi}\\
&\qquad+\log\Big(1-2\cos(2\pi l)\hat{Y}(\theta-i\phi_{r+1}+\frac{i\pi}{2})+\hat{Y}(\theta-i\phi_{r+1}+\frac{i\pi}{2})^{2}\Big),\\
\log\hat{Y}(\theta-i\hat{\phi})&=-|\hat{m}|e^{\theta}+\int_{\mathbb{R}}\frac{\log\Big(1+Y_{r+1}(\theta^{\prime}-i\hat{\phi})\Big)}{\cosh(\theta-\theta^{\prime}-i\hat{\phi}+i\phi_{r+1})}\frac{d\theta^{\prime}}{2\pi}
+\log\Big(1+Y_{r+1}(\theta-i\hat{\phi}-\frac{i\pi}{2})\Big).
\ea
\ee
Note that $Y_{r+1}(\theta-i\hat{\phi}-\frac{i\pi}{2})$ and $\hat{Y}(\theta-i\phi_{r+1}+\frac{i\pi}{2})$ do not appear on the left hand side. To obtain a closed TBA system, we evaluate the TBA equations at $\theta+i\phi_{r+1}-i\hat{\phi}-\frac{i\pi}{2}$ and $\theta+i\hat{\phi}-i\phi_{r+1}+\frac{i\pi}{2}$, and obtain
\be
\ba
\log Y_{r+1}(\theta-i\hat{\phi}-\frac{i\pi}{2})&=-|m_{r+1}|e^{\theta+i\phi_{r+1}-i\hat{\phi}-\frac{i\pi}{2}}+\int_{\mathbb{R}}\frac{\log\Big(1-2\cos(2\pi l)\hat{Y}(\theta^{\prime}-\hat{\phi})+\hat{Y}(\theta^{\prime}-\hat{\phi})^{2}\Big)}{\cosh(\theta-\theta^{\prime}-\frac{i\pi}{2})}\frac{d\theta^{\prime}}{2\pi}\\
&\qquad+\log\Big(1-2\cos(2\pi l)\hat{Y}(\theta-i\hat{\phi})+\hat{Y}(\theta-i\hat{\phi})^{2}\Big),\\
\log\hat{Y}(\theta-i\phi_{r+1}+\frac{i\pi}{2})&=-|\hat{m}|e^{\theta+i\hat{\phi}-i\phi_{r+1}+\frac{i\pi}{2}}+\int_{\mathbb{R}}\frac{\log\Big(1+Y_{r+1}(\theta^{\prime}-i{\phi}_{r+1})\Big)}{\cosh(\theta-\theta^{\prime}+\frac{i\pi}{2})}\frac{d\theta^{\prime}}{2\pi}\\
&\qquad+\log\Big(1+Y_{r+1}(\theta-i\phi_{r+1})\Big).
\ea
\ee
These two equations together with (\ref{eq:TBA-WC-Yr+1-Yhat}) and the TBA equations with $s=1,\cdots,r$ in (\ref{eq:TBA-com-mass}) form a closed TBA system.


\section{Duality between ODEs}
In this appendix, we discuss the relations among the present ODE and other types of ODEs under a change of variable $x\to \tilde{x}$.
This relation provides a further consistency check of the correspondence between the ODEs and the TBA systems.
The Schr\"odinger type equation (\ref{eq:Schrodinger-xu}) is related with the Schr\"odinger type equation studied in \cite{Ito:2018eon} by coordinate transformation $x\to \tilde{x}$. Then it is natural to ask how the TBA equations (\ref{eq:TBA-pole}) are related with the TBA equations found in \cite{Ito:2018eon}. More general, one may consider the transformation from the ODE in $x$ $({\rm ODE}_{x})$ to the ODE in $\tilde{x}$ $(\widetilde{\rm ODE}_{\tilde{x}})$, and ask how their TBA equations transform in this procedure. This may help us to find the TBA equations for the ODE with more general potential. 
In this appendix, we start with certain Schr\"odinger-type equation with polynomial potential studied in \cite{Ito:2018eon} and their TBA equations in the minimal chamber. We then impose the coordinate transformation to get the Schr\"odinger type equation (\ref{eq:Schrodinger-xu}). By comparing their TBA equations, we show how they are related to each other.

Let us begin with the Schr\"odinger equation with potential of even power polynomial in $x$, which has been studied in \cite{Ito:2018eon}
\be
\label{eq:Schrodinger-2n}
\Big(-\zeta^{2}\frac{d^{2}}{dx^{2}}+x^{2n}+u_{1}x^{2(n-1)}+u_{2}x^{2(n-2)}+\cdots+u_{n}\Big)\psi=0
\ee
with a positive integer $n>1$. We assume that all the turning points are real and ordered as
\be
-a_1<-a_2\cdots<-a_n<a_n<\cdots<a_2<a_1.
\ee
In this case we obtain the $A_{2n-1}/Z_2$-type TBA equations
\be
\ba
\label{eq:n-TBA}
\log Y_{1}(\theta)&=-m_{1}e^{\theta}+\int_{\mathbb{R}}\frac{\log\Big(1+Y_{2}(\theta^{\prime})\Big)}{\cosh(\theta-\theta^{\prime})}\frac{d\theta^{\prime}}{2\pi},
\\
\log Y_{2}(\theta)&=-m_{2}e^{\theta}+\int_{\mathbb{R}}\frac{\log\Big(1+Y_{1}(\theta^{\prime})\Big)}{\cosh(\theta-\theta^{\prime})}\frac{d\theta^{\prime}}{2\pi}+\int_{\mathbb{R}}\frac{\log\Big(1+Y_{3}(\theta^{\prime})\Big)}{\cosh(\theta-\theta^{\prime})}\frac{d\theta^{\prime}}{2\pi},
\\&
\vdots,\\
\log Y_{n}(\theta)&=-m_{n}e^{\theta}+2\int_{\mathbb{R}}\frac{\log\Big(1+Y_{n-1}(\theta^{\prime})\Big)}{\cosh(\theta-\theta^{\prime})}\frac{d\theta^{\prime}}{2\pi}.
\ea
\ee
The masses are given by
\be
\ba
m_{2k+1}&=\frac{2}{i}\int_{a_{2k+1}}^{a_{2k+2}}\sqrt{x^{2n}+u_{1}x^{2(n-1)}+\cdots+u_{n}}dx,
\\
m_{2k}&=2\int_{a_{2k}}^{a_{2k+1}}\sqrt{x^{2n}+u_{1}x^{2(n-1)}+\cdots+u_{n}}dx,
\ea
\ee
where $1\leq 2k, 2k+1\leq n$. 

Setting $x^2=\tilde{x}$, the Schr\"odinger equation (\ref{eq:Schrodinger-2n}) becomes
\be
\label{eq:Schrodinger-n+pole}
\Big(-\zeta^{2}\frac{d^{2}}{d\tilde{x}^{2}}+\frac{1}{4}\tilde{x}^{n-1}+\frac{u_{1}}{4}\tilde{x}^{n-2}+\frac{u_{2}}{4}\tilde{x}^{n-3}+\cdots+\frac{u_{n}}{4\tilde{x}}-\zeta^{2}\frac{3}{16\tilde{x}^{2}}\Big)\tilde{\psi}(\tilde{x})=0,
\ee
where $\tilde{\psi}(\tilde{x})=e^{-\frac{1}{2}\int\frac{1}{2\tilde{x}}d\tilde{x}}\psi(x)$. Note that this Schr\"odinger equation has the form of (\ref{eq:Schrodinger-xu}) but with $l=-\frac{1}{4}$ or $-\frac{3}{4}$. The turning points are
\be
a_n^2<\cdots<a_2^2<a_1^2.
\ee
The TBA equations are
\be
\ba
\label{eq:TBA-ct}
\log\tilde{Y}_{1}(\theta)&=-\tilde{m}_{1}e^{\theta}+\int_{\mathbb{R}}\frac{\log\Big(1+\tilde{Y}_{2}(\theta^{\prime})\Big)}{\cosh(\theta-\theta^{\prime})}\frac{d\theta^{\prime}}{2\pi},\\
\log\tilde{Y}_{2}(\theta)&=-\tilde{m}_{2}e^{\theta}+\int_{\mathbb{R}}\frac{\log\Big(1+\tilde{Y}_{1}(\theta^{\prime})\Big)}{\cosh(\theta-\theta^{\prime})}\frac{d\theta^{\prime}}{2\pi}+\int_{\mathbb{R}}\frac{\log\Big(1+\tilde{Y}_{3}(\theta^{\prime})\Big)}{\cosh(\theta-\theta^{\prime})}\frac{d\theta^{\prime}}{2\pi},
\\&
\vdots ,\\\log\tilde{Y}_{n-1}(\theta)&=-\tilde{m}_{n-1}e^{\theta}+\int_{\mathbb{R}}\frac{\log\Big(1+\tilde{Y}_{n-2}(\theta^{\prime})\Big)}{\cosh(\theta-\theta^{\prime})}\frac{d\theta^{\prime}}{2\pi}+\int_{\mathbb{R}}\frac{\log\Big(1+\tilde{\hat{Y}}(\theta^{\prime})^{2}\Big)}{\cosh(\theta-\theta^{\prime})}\frac{d\theta^{\prime}}{2\pi},\\
\log\tilde{\hat{Y}}(\theta)&=-\tilde{\hat{m}}e^{\theta}+\int_{\mathbb{R}}\frac{\log\Big(1+\tilde{Y}_{2}(\theta^{\prime})\Big)}{\cosh(\theta-\theta^{\prime})}\frac{d\theta^{\prime}}{2\pi},
\ea
\ee
whose masses are given by
\be
\ba
\tilde{m}_{2k+1}&=\frac{2}{i}\int_{a_{2k+2}^{2}}^{a_{2k+1}^{2}}\sqrt{\frac{\tilde{x}^{n}+u_{1}\tilde{x}^{n-1}+u_{2}\tilde{x}^{n-2}+\cdots+u_{n}}{4\tilde{x}}}d\tilde{x},\\
\tilde{m}_{2k}&=2\int_{a_{2k+1}^{2}}^{a_{2k}^{2}}\sqrt{\frac{\tilde{x}^{n}+u_{1}\tilde{x}^{n-1}+u_{2}\tilde{x}^{n-2}+\cdots+u_{n}}{4\tilde{x}}}d\tilde{x},\quad 1\leq 2k,2k+1\leq n-1,\\
\tilde{\hat{m}}&=\begin{cases}
\frac{2}{i}\int_{0}^{a_{n}}\sqrt{\big(x^{2n}+u_{1}x^{2(n-1)}+u_{2}x^{2(n-2)}+\cdots+u_{n}\big)} & n:{\rm odd}\\
2\int_{0}^{a_{n}}\sqrt{\big(x^{2n}+u_{1}x^{2(n-1)}+u_{2}x^{2(n-2)}+\cdots+u_{n}\big)} & n:{\rm even}
\end{cases}.
\ea
\ee
After simple calculations, we find
\be
\tilde{m}_{a}=m_a,\quad \tilde{\hat{m}}=\frac{1}{2}m_n\quad\mbox{for}\quad a=1,\cdots,n-1.
\ee
Comparing with the TBA equations (\ref{eq:n-TBA}) and (\ref{eq:TBA-ct}), we find
\be
\tilde{Y}_1(\theta)=Y_1(\theta),\quad \tilde{Y}_2(\theta)=Y_2(\theta),\cdots, \tilde{Y}_{n-1}(\theta)=Y_{n-1}(\theta),\quad \tilde{\hat{Y}}(\theta)^2=Y_n(\theta).
\ee
Therefore, the two sets of TBA equation derived from (\ref{eq:Schrodinger-2n}) and (\ref{eq:Schrodinger-n+pole}) respectively are related with each other by redefinition.


\end{document}